\documentclass{nature3}
\usepackage{graphicx}
\usepackage{xcolor}
\usepackage{float}
\usepackage{longtable}
\usepackage{tabularx}
\usepackage{url}
\usepackage{amsmath}
\usepackage{amssymb}
\usepackage{subcaption}
\usepackage{ulem}
\usepackage{courier}
\usepackage[labelfont=bf, labelsep=none]{caption}
\usepackage[super]{natbib}
\usepackage[colorlinks=true,citecolor=blue,urlcolor=cyan]{hyperref}
\usepackage{chemformula}
\usepackage{gensymb}

\usepackage[affil-it]{authblk}

\newcommand{\pyratbay}{\texttt{Pyrat Bay}}  

\newcommand{\sep}{$\,\vert\,$}

\newcommand{\arxiv}{y}

\ifthenelse{\equal{\arxiv}{n}}{
\usepackage{xr}
\externaldocument{supplementary-nature}
}{}

\DeclareSymbolFont{UPM}{U}{eur}{m}{n}
\DeclareMathSymbol{\umu}{0}{UPM}{"16}
\newcommand\micron{\ifmmode{\umu\rm m}\else{$\umu$m}\fi}
\newcommand\microns{\micron}

\usepackage{lineno}
\makeatletter
\def\@maketitle{%
  \newpage
  \begin{center}%
  \let \footnote \thanks
    {\Large \@title \par}%
    \vskip -3em
    {\large
      \begin{tabular}[t]{c}%
        \@author
      \end{tabular}\par}%
  \end{center}%
  \par
  \vskip 0.5em}
\makeatother

\title{Horizontal and vertical exoplanet thermal structure from a JWST spectroscopic eclipse map}

\author[1,2,*]{Ryan C. Challener}

\newcounter{ind}

\affil[*]{Co-First Author and Corresponding Author}
\stepcounter{ind}\edef\Cornell{\arabic{ind}}
\affil[\arabic{ind}]{Department of Astronomy, Cornell University, Ithaca, NY, USA}
\stepcounter{ind}\edef\Umich{\arabic{ind}}
\affil[\arabic{ind}]{Department of Astronomy, University of Michigan, Ann Arbor, MI, USA}

\stepcounter{ind}\edef\ASU{\arabic{ind}}
\affil[\arabic{ind}]{The School of Earth and Space Exploration, Arizona State University, Tempe, AZ, USA}
\stepcounter{ind}\edef\Maryland{\arabic{ind}}
\affil[\arabic{ind}]{Department of Astronomy, University of Maryland, College Park, MD, USA}
\stepcounter{ind}\edef\Pegasi{\arabic{ind}}
\affil[\arabic{ind}]{51 Pegasi b Fellow}
\author[\ASU,\Maryland,\Pegasi,*]{Megan Weiner Mansfield} 

\stepcounter{ind}\edef\SRI{\arabic{ind}}
\affil[\arabic{ind}]{Space Research Institute, Austrian Academy of Sciences, Graz, Austria}
\stepcounter{ind}\edef\INAF{\arabic{ind}}
\affil[\arabic{ind}]{INAF - Turin Astrophysical Observatory, Pino Torinese, Italy}
\author[\SRI,\INAF]{Patricio E. Cubillos}

\stepcounter{ind}\edef\EPL{\arabic{ind}}
\affil[\arabic{ind}]{Earth and Planets Laboratory, Carnegie Institution for Science, Washington, DC, USA}
\author[\EPL]{Anjali A. A. Piette}

\stepcounter{ind}\edef\Montreal{\arabic{ind}}
\affil[\arabic{ind}]{Department of Physics and Institute for Research on Exoplanets, Université de Montréal, Montreal, QC, Canada}
\author[\Montreal]{Louis-Philippe Coulombe}

\author[\Maryland]{Hayley Beltz}

\stepcounter{ind}\edef\NYUAD{\arabic{ind}}
\affil[\arabic{ind}]{Department of Physics, New York University Abu Dhabi, Abu Dhabi, UAE}
\stepcounter{ind}\edef\CASS{\arabic{ind}}
\affil[\arabic{ind}]{Center for Astrophysics and Space Science (CASS), New York University Abu Dhabi, Abu Dhabi, UAE}
\author[\NYUAD,\CASS]{Jasmina Blecic}

\author[\Umich]{Emily Rauscher}

\stepcounter{ind}\edef\Chicago{\arabic{ind}}
\affil[\arabic{ind}]{Department of Astronomy \& Astrophysics, University of Chicago, Chicago, IL, USA}
\author[\Chicago]{Jacob L. Bean}

\stepcounter{ind}\edef\UCLA{\arabic{ind}}
\affil[\arabic{ind}]{Department of Earth, Planetary, and Space Sciences, University of California, Los Angeles, CA USA}
\stepcounter{ind}\edef\DPTrottier{\arabic{ind}}
\affil[\arabic{ind}]{Department of Physics and Trottier Institute for Research on Exoplanets, Universit\'{e} de Montr\'{e}al, Montreal, QC, Canada}

\author[\UCLA,\DPTrottier]{Bj\"orn Benneke}

\author[\Maryland]{Eliza M.-R. Kempton}

\stepcounter{ind}\edef\UCF{\arabic{ind}}
\affil[\arabic{ind}]{Planetary Sciences Group, Department of Physics and Florida Space Institute, University of Central Florida, Orlando, Florida, USA}
\author[\UCF]{Joseph Harrington}

\author[\Maryland]{Thaddeus D. Komacek}

\stepcounter{ind}\edef\Nice{\arabic{ind}}
\affil[\arabic{ind}]{Université Côte d'Azur, Observatoire de la Côte d'Azur, CNRS, Laboratoire Lagrange, France}
\author[\Nice]{Vivien Parmentier}

\stepcounter{ind}\edef\Leicester{\arabic{ind}}
\affil[\arabic{ind}]{School of Physics and Astronomy, University of Leicester, Leicester}
\author[\Leicester]{S. L. Casewell}

\stepcounter{ind}\edef\DLR{\arabic{ind}}
\affil[\arabic{ind}]{Institute of Planetary Research, German Aerospace Center (DLR), Berlin, Germany}
\author[\DLR]{Nicolas Iro}

\stepcounter{ind}\edef\Rome{\arabic{ind}}
\affil[\arabic{ind}]{Department of Physics, University of Rome ``Tor Vergata'', Rome, Italy}
\stepcounter{ind}\edef\MPI{\arabic{ind}}
\affil[\arabic{ind}]{Max Planck Institute for Astronomy, Heidelberg, Germany}
\author[\Rome, \MPI, \INAF]{Luigi Mancini}

\author[\Maryland]{Matthew C. Nixon}

\stepcounter{ind}\edef\Trottier{\arabic{ind}}
\affil[\arabic{ind}]{Trottier Institute for Research on Exoplanets, University of Montréal, Montréal, Québec, Canada}
\author[\Trottier]{Michael Radica}

\author[\Chicago,\Pegasi]{Maria E. Steinrueck}

\stepcounter{ind}\edef\Sagan{\arabic{ind}}
\affil[\arabic{ind}]{NHFP Sagan Fellow}
\author[\ASU,\Sagan]{Luis Welbanks}

\stepcounter{ind}\edef\SantaCruzDAA{\arabic{ind}}
\affil[\arabic{ind}]{Department of Astronomy \& Astrophysics, University of California Santa Cruz, Santa Cruz, California, USA}
\author[\SantaCruzDAA]{Natalie M. Batalha}

\stepcounter{ind}\edef\CATA{\arabic{ind}}
\affil[\arabic{ind}]{Instituto de Astrofisica, Departamento de Ciencias Fisicas, Facultad de Ciencias Exactas, Universidad Andres Bello, Santiago, Chile}
\author[\CATA]{Claudio Caceres}

\stepcounter{ind}\edef\Kansas{\arabic{ind}}
\affil[\arabic{ind}]{Department of Physics \& Astronomy, University of Kansas, Lawrence, KS, USA}
\author[\Kansas]{Ian J.M. Crossfield}

\stepcounter{ind}\edef\Kapteyn{\arabic{ind}}
\affil[\arabic{ind}]{Kapteyn Astronomical Institute, Rijksuniversiteit Groningen, Groningen, The Netherlands}
\author[\Kapteyn]{Nicolas Crouzet}

\stepcounter{ind}\edef\Amsterdam{\arabic{ind}}
\affil[\arabic{ind}]{Anton Pannekoek Institute for Astronomy, University of Amsterdam, Amsterdam, The Netherlands}
\author[\Amsterdam]{Jean-Michel D\'esert}

\stepcounter{ind}\edef\Munchen{\arabic{ind}}
\affil[\arabic{ind}]{Universitäts-Sternwarte, Ludwig-Maximilians-Universität München, München, Germany}
\stepcounter{ind}\edef\Garching{\arabic{ind}}
\affil[\arabic{ind}]{Exzellenzcluster Origins, Garching, Germany}
\author[\Munchen,\Garching]{Karan Molaverdikhani}

\stepcounter{ind}\edef\Stsci{\arabic{ind}}
\affil[\arabic{ind}]{Space Telescope Science Institute, Baltimore, MD, USA}
\author[\Stsci]{Nikolay K. Nikolov}

\stepcounter{ind}\edef\IAC{\arabic{ind}}
\affil[\arabic{ind}]{Instituto de Astrofísica de Canarias (IAC), Tenerife, Spain}
\author[\IAC]{Enric Palle}

\stepcounter{ind}\edef\DEAPMIT{\arabic{ind}}
\affil[\arabic{ind}]{Department of Earth, Atmospheric and Planetary Sciences, Massachusetts Institute of Technology, Cambridge, MA, USA}
\stepcounter{ind}\edef\KavliMIT{\arabic{ind}}
\affil[\arabic{ind}]{Kavli Institute for Astrophysics and Space Research, Massachusetts Institute of Technology, Cambridge, MA, USA}
\author[\DEAPMIT,\KavliMIT]{Benjamin V. Rackham}

\stepcounter{ind}\edef\Steward{\arabic{ind}}
\affil[\arabic{ind}]{Steward Observatory, University of Arizona, Tucson, AZ, USA}
\author[\Steward]{Everett Schlawin}

\stepcounter{ind}\edef\JHUDEPS{\arabic{ind}}
\affil[\arabic{ind}]{Department of Earth and Planetary Sciences, Johns Hopkins University, Baltimore, MD, USA}
\stepcounter{ind}\edef\JHUDPA{\arabic{ind}}
\affil[\arabic{ind}]{Department of Physics \& Astronomy, Johns Hopkins University, Baltimore, MD, USA}
\author[\JHUDEPS,\JHUDPA]{David K. Sing}

\stepcounter{ind}\edef\APL{\arabic{ind}}
\affil[\arabic{ind}]{Johns Hopkins APL, Laurel, MD, USA}
\author[\APL]{Kevin B. Stevenson}

\stepcounter{ind}\edef\TLI{\arabic{ind}}
\affil[\arabic{ind}]{Tsung-Dao Lee Institute, Shanghai Jiao Tong University, 520 Shengrong Road, Shanghai, 200127, People’s Republic of China}
\author[\TLI]{Xianyu Tan}

\author[\Cornell,\Sagan]{Jake D. Turner}

\stepcounter{ind}\edef\SantaCruzDEPS{\arabic{ind}}
\affil[\arabic{ind}]{Department of Earth and Planetary Sciences, University of California Santa Cruz, Santa Cruz, California, USA}
\author[\SantaCruzDEPS]{Xi Zhang}

\DeclareCaptionFormat{cont}{#1 (cont.)#2#3\par}

\begin{document}

\maketitle

\noindent
Corresponding Author Contacts: rcc276@cornell.edu (Ryan C. Challener) and mwm@umd.edu (Megan Weiner Mansfield)\\

\begin{abstract}
Highly-irradiated giant exoplanets known as ``ultra-hot Jupiters'' are anticipated to exhibit large variations of atmospheric temperature and chemistry as a function of longitude, latitude, and altitude. Previous observations have hinted at these variations, but the existing data have been fundamentally restricted to probing hemisphere-integrated spectra, thereby providing only coarse information on atmospheric gradients. Here we present a spectroscopic eclipse map of an extrasolar planet, resolving the atmosphere in multiple dimensions simultaneously. We analyze a secondary eclipse of the ultra-hot Jupiter WASP-18b observed with the NIRISS instrument on JWST. The mapping reveals weaker longitudinal temperature gradients than were predicted by theoretical models, indicating the importance of hydrogen dissociation and/or nightside clouds in shaping global thermal emission. Additionally, we identify two thermally distinct regions of the planet's atmosphere: a “hotspot” surrounding the substellar point and a “ring” near the dayside limbs. The hotspot region shows a strongly inverted thermal structure due to the presence of optical absorbers and a water abundance marginally lower than the hemispheric average, in accordance with theoretical predictions. The ring region shows colder temperatures and poorly constrained chemical abundances. Similar future analyses will reveal three-dimensional thermal, chemical, and dynamical properties of a broad range of exoplanet atmospheres.

\end{abstract}

%#####################################################################################################################

%intro

%Observations
\section{Introduction}

As part of the JWST Early Release Science Program\cite{Bean2018}, we observed a secondary eclipse of WASP-18b with the first order of the NIRISS Single-object Slitless Spectroscopy (SOSS) mode\cite{AlbertEtal2023paspSOSS} covering 0.85 -- 2.85 $\mu$m. The dayside spectrum of the planet revealed an inverted vertical temperature profile, the presence of water in the atmosphere, and evidence for short-wavelength absorbers such as H$^-$, TiO, or VO \cite{Coulombe2023}. A broadband eclipse map of the planet showed the planet's hottest hemisphere is aligned with the substellar point and there are steep temperature gradients from the substellar point to the limbs\cite{Coulombe2023}, both indicators for atmospheric drag\cite{Arcangeli2019, Beltz2022}.

% Ryan's attempt at some text
% Looks good to me, adding it with some minor edits -MM

Here, we %combine the 2D (horizontal) information from flux vs.\ time at a given wavelength with the vertical information from spectroscopy 
reanalyze the NIRISS data by applying the secondary eclipse mapping method at multiple wavelengths to infer the multidimensional temperature structure of WASP-18b's atmosphere\cite{Cowan2018}. %I think this sentence is a bit confusing but I'm still trying to figure out how to better say it -MM
We used the \texttt{Eigenspectra}\cite{Mansfield2020} method to reanalyze the wavelength-resolved, systematics-corrected light curves presented in Coulombe et al. (2023)\cite{Coulombe2023}.
Briefly, we fit a 2D brightness map independently at each wavelength to the eclipse ingress, egress, and out-of-eclipse phase variation\cite{Rauscher2018}. As described in the Methods, these fits were strongly preferred over simple sinusoid fits to the out-of-eclipse phase variation in more than half of the wavelength bins, with a Bayesian Information Criterion (BIC) difference of $\Delta \mathrm{BIC} \geq 10$ in favor of the \texttt{Eigenspectra} model.
Then, we stacked the individual wavelength maps together, identified spatial regions of the planet (groups) that are spectroscopically similar, and extracted the spectra from each group. 
These spectra were then analyzed with traditional 1D characterization approaches to determine the vertical temperature structures and chemical compositions of each region.
In this work, we used \texttt{HyDRA}\cite{Gandhi2018, Gandhi2020, Piette2020, Piette2022} and {\pyratbay}\cite{CubillosBlecic2021mnrasPyratBay} to make atmospheric inferences (i.e., atmospheric retrieval).
%A thorough description of \texttt{Eigenspectra} and a second method we use to cross-check our results can be found in the Methods section.
Thorough descriptions of \texttt{Eigenspectra}, the retrievals on \texttt{Eigenspectra}, and a second eclipse mapping method (\texttt{ThERESA}) we use to cross-check our results can be found in the Methods section.

\section{Results}

\subsection{Wavelength-dependent Maps}

Because of the small size of eclipse-mapping signals and the computational intensity of mapping fits, we performed eclipse mapping on 25 light curves binned down evenly in wavelength from the higher-resolution light curves fitted in Ref. \citep{Coulombe2023}. We also tested a lower-resolution spectrum and found similar results (see Methods). However, we note that generally the effect of wavelength resolution on spectroscopic eclipse mapping has not been studied in detail and should be examined in future work. Figure \ifthenelse{\equal{\arxiv}{y}}{\ref{fig:maps_8bin}}{1} shows the 2D brightness temperature maps for each of the 25 wavelengths, and Extended Data Figure \ifthenelse{\equal{\arxiv}{y}}{\ref{fig:lightcurve25}}{1} shows the \texttt{Eigenspectra} fits to each of the 25 bins.

Before stacking the single-wavelength maps, we constructed longitudinal brightness profiles by weighting the retrieved 2D maps from \texttt{Eigenspectra} by the squared cosine of the latitude. Figure \ifthenelse{\equal{\arxiv}{y}}{\ref{fig:longprof}}{2} shows the longitudinal brightness profiles for \texttt{Eigenspectra}. Uncertainties on the longitudinal profiles from \texttt{Eigenspectra} are low ($\approx$5-10\%) because the data are well fit by only two or three non-uniform map components, depending on wavelength, limiting model flexibility to large-scale variations. Hotspot offsets from \texttt{Eigenspectra} ranged from $-5\degree$ to $7\degree$, with uncertainties of $\sim1\degree$ (Supplementary Table \ref{tab:eigenspec25}). This trend of small to negligible hotspot offsets for all wavelengths examined is in agreement with the previous analysis of the full dayside observations \citep{Coulombe2023}.

Figure \ifthenelse{\equal{\arxiv}{y}}{\ref{fig:longprof}}{2} also compares the retrieved longitudinal profiles to predictions from two general circulation models (GCMs) previously compared to the full secondary eclipse spectrum\cite{Coulombe2023}: the SPARC/MITgcm\cite{ShowmanEtal2009apjSPARC}, which includes a uniform drag, and the RM-GCM\cite{Beltz2022}, which uses a kinematic magnetohydrodynamic (MHD) drag. Both GCMs were computed assuming solar atmospheric metallicity and carbon-to-oxygen ratio. We note that, at nearly all wavelengths, the RM-GCM is brighter than the SPARC/MITgcm, which is likely due to different radiative transfer methods resulting in different thermal structures between the two GCMs. The observed lack of significant hotspot offsets agrees with the GCMs, as expected based on predictions that ultra-hot Jupiters will experience increased magnetic atmospheric drag slowing the planet's equatorial jet\citep{Tan2019,Beltz2022}. More complex treatments of MHD additionally predict oscillating hotspot offsets due to non-linear Reynolds stresses \citep{Rogers2017, Hindle2021b}, which could be confirmed with follow-up observations at different epochs.

%Because the photospheric pressure changes with wavelength, 
Spectroscopic eclipse mapping observations are in theory able to probe the temperature structure and hotspot offset over a range of pressures because the photospheric pressure changes with wavelength. For hot Jupiters such as WASP-18b, GCMs generally predict increasing hotspot offsets at deeper pressures\cite{Komacek2022,Beltz2022}. We searched for trends in the hotspot offset as a function of retrieved pressure but found no clear trends. %This may be due to the broad range of pressures probed by each relatively wide wavelength band, {\color{red}will need to change this} although we note that the higher-resolution wavelength binning did not produce clearer results. It also 
This may be due to the difficulty of seeing trends in hotspot offset for WASP-18b in particular, which as one of the hottest known ultra-hot Jupiters is predicted to have smaller hotspot offsets than slightly cooler planets throughout much of its atmosphere \cite{Showman2020}. Future work to apply spectroscopic eclipse mapping to cooler planets, which are expected to have more variation in hotspot position, may reveal stronger trends. 
 
 Notably, the GCMs also predict a steeper decrease in flux away from the substellar point than the \texttt{Eigenspectra} maps at most wavelengths, leading to cooler predicted temperatures near the limb at $\pm90\degree$ longitude. GCMs with weaker drag cannot account for this difference, as they all show larger hotspot offsets than what is seen in the \texttt{Eigenspectra} longitudinal profiles. The warmer-than-predicted limbs may be due to the influence of hydrogen dissociation and recombination, which increases day-night heat transport\cite{Tan2019,Mansfield2020}. We tested models including H$_{2}$ dissociation and a uniform drag and found no significant warming near the limbs. However, future models combining H$_{2}$ dissociation and a non-uniform drag may result in more significant limb warming. Alternatively, the warm limbs could be indicative of nightside clouds, which would warm the atmosphere and potentially change the substellar-point-to-limb temperature gradient\cite{Roman2019,Parmentier2021}, but are not included in the GCMs shown here. %and could This difference likely indicates the influence of hydrogen dissociation on the rate of temperature change across longitudes. Dayside hydrogen dissociation and nightside recombination, which is not incorporated into the GCMs shown here, acts to increase day-night heat transport, decreasing the sharpness of the day-night temperature difference\cite{Tan2019,Mansfield2020}, similar to what is seen in the longitudinal profiles.

%{\color{red}FINDME: End of rewrite}
\subsection{Horizontal and Vertical Map}
We then applied the full \texttt{Eigenspectra} method to identify spatial regions with similarly shaped spectra. As shown in Figure \ifthenelse{\equal{\arxiv}{y}}{\ref{fig:maps_8bin}}{1}, the \texttt{Eigenspectra} mapping method identified three regions of the map with distinct spectral shapes, which are roughly concentric circles centered on the substellar point.
This demonstrates the multidimensional information in these data, and the need for a multidimensional approach to interpreting them: if a uniform planet (i.e., one with even no phase variation out of eclipse) was a reasonable assumption within our data precision, \texttt{Eigenspectra} would find only one distinct region in the planet. We note that these three groups are a discrete approximation of a planet that likely has continuously varying properties, which we discuss further in the Methods.
We refer to these three groups as the ``hotspot'', ``ring'', and ``outer'' groups, in order of their angular distance from the substellar point. 
The outer group, shown in Supplementary Figure \ifthenelse{\equal{\arxiv}{y}}{\ref{fig:eigen_spectra_check}}{2}, had a %very low 
signal-to-noise ratio about $2-12\times$ lower than the other groups because it contains regions of the planet only observed very briefly near the beginning or end of the observation as part of the nightside rotated into view (Supplementary Table \ifthenelse{\equal{\arxiv}{y}}{\ref{tab:eigenspectra}}{1}). %Additionally, its shape may be influenced by the fitting method, as discussed in more detail in the Methods. 
Therefore, we limit our analysis to the hotspot and ring groups. 
Figure \ifthenelse{\equal{\arxiv}{y}}{\ref{fig:spectra}}{3} shows the emission spectra from the hotspot and ring groups, along with 1D best-fitting models to those spectra.
These spectra, as expected, bracket the hemispherically-averaged dayside spectrum from ref. \cite{Coulombe2023}, with the hotspot spectrum $\sim150$ K hotter and the ring spectrum $\sim400$ K colder.
Indeed, an average of the flux emitted from these regions, when appropriately accounting for viewing geometry and relative area, closely matches the dayside average emission spectrum (Supplementary Figure \ifthenelse{\equal{\arxiv}{y}}{\ref{fig:eigen_spectra_check}}{2}).

\section{Discussion}
\begin{figure}
    \centering
    \ifthenelse{\equal{\arxiv}{y}}{
    \includegraphics[width=\linewidth]{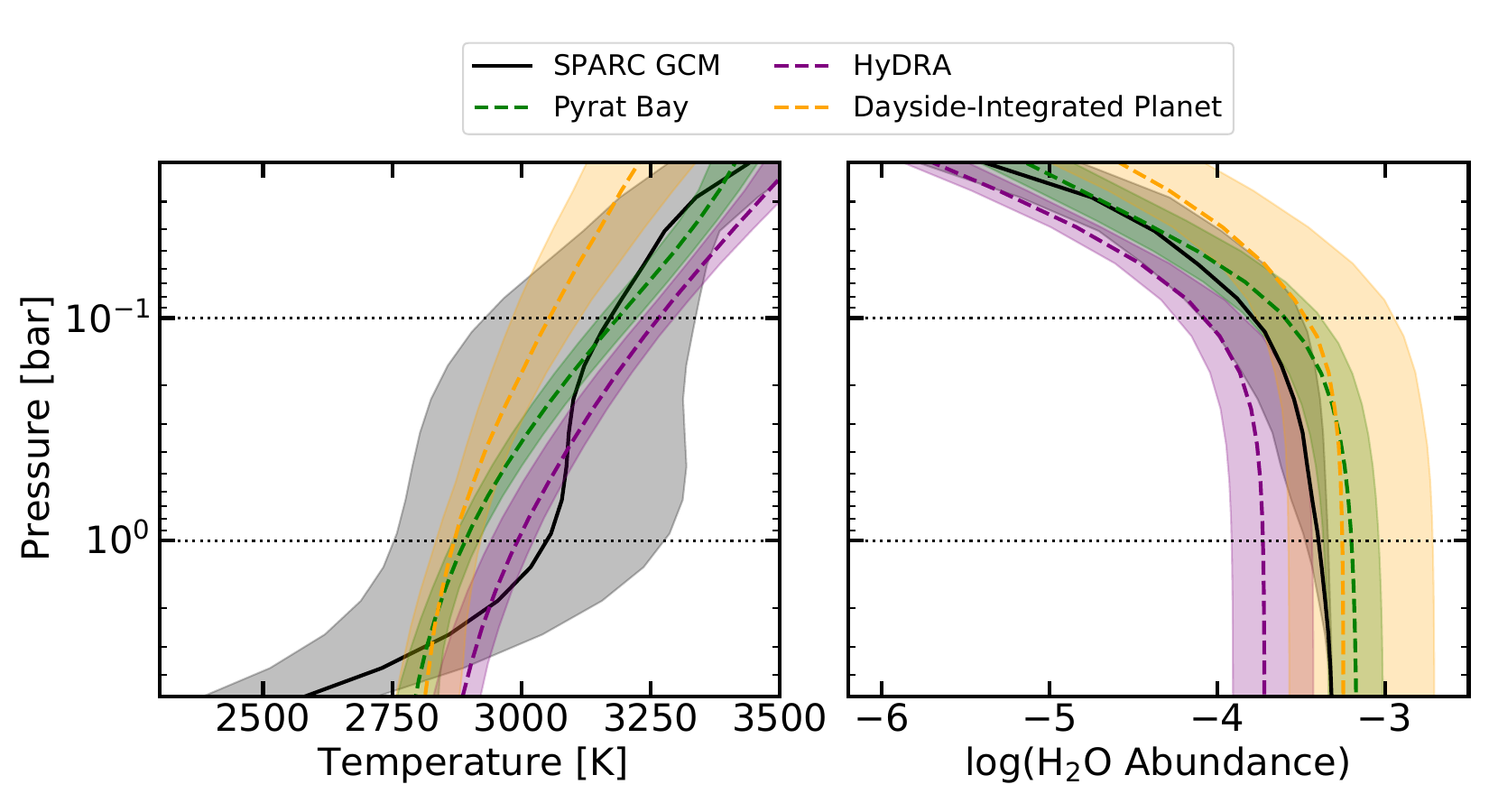}
    }{}
    \caption{\sep The \texttt{Eigenspectra} hotspot group shows a slightly hotter temperature and slightly lower water abundance than the full dayside-integrated result. %Leftmost and rightmost plots show T-P profiles for the ``hotspot'' and ``ring'' groups, respectively, while the middle plots show H$_{2}$O and H$^{-}$ abundances for the ``hotspot'' group. 
    The left plot shows T-P profiles, and the right plot shows retrieved  H$_{2}$O abundances. In both plots, purple and green lines show retrievals on the \texttt{Eigenspectra} hotspot spectrum using \texttt{HyDRA} and {\pyratbay}, respectively, and yellow lines show the retrieval on the full dayside spectrum from Ref.\cite{Coulombe2023}, with shading indicating 68.3\% credible region ($1\sigma$) uncertainties. Black lines show average profiles in the group region from a SPARC/MIT GCM of solar atmosphere metallicity and carbon-to-oxygen ratio, and black-shaded regions show the full range of per-point GCM profiles in the region of the hotspot group. Black dotted lines indicate the approximate range of pressures that our observation probes. Note that the full planet retrieval was also performed with \texttt{HyDRA}, and so is more similar in retrieval set-up to the \texttt{HyDRA} hotspot retrieval than that of {\pyratbay}. For reference, the equilibrium temperature of WASP-18b is $\approx2400$~K\cite{CortesZuleta2020}.%Abundances are not shown for the ``ring'' group because the retrievals for that group did not produce good constraints on abundances.
    }
    \label{fig:retrieved_abundances_mainfig}
\end{figure}

Atmospheric inference of the hotspot group shows a thermal structure and composition that is consistent with the same approach applied to the full dayside spectrum, and similar to expectations from GCMs (Figure \ifthenelse{\equal{\arxiv}{y}}{\ref{fig:retrieved_abundances_mainfig}}{4}).
The hotspot thermal profile is marginally hotter than the dayside average but shows a similar thermal inversion (increasing temperature with decreasing pressure) at the planet's near-infrared photosphere, which is likewise predicted by GCMs.
The retrieved H$_2$O abundances are consistent with the GCM but marginally lower than the full-dayside retrieval, likely due to factors such as increased thermal dissociation of H$_{2}$O in the hotter hotspot group. %the higher temperature of the ``hotspot'' group resulting in more H$_2$O dissociation in that region.
%The retrieved H$^-$ abundances are consistent with the full dayside retrieval but are much more constrained.
While the retrievals do not tightly constrain any other individual chemical abundances, we find evidence for optical opacity sources (a combination of H$^{-}$/TiO/VO), likely drivers of the thermal inversion, at $5.1\sigma$.
Evidence for optical opacity is also seen in the spatially averaged dayside spectrum (a weighted average of the hotspot, ring and outer spectra), though with a slightly lower detection significance of $4.6\sigma$. Such similarities are expected, as the hotspot region is both bright and directly visible throughout the observation, and therefore dominates the planet's dayside emission. However, changes in chemistry and vertical temperature gradient away from the hotspot may weaken the detection of optical species in the averaged dayside spectrum. Additionally, while the hotspot thermal structure and water abundance are not significantly different from those of the full dayside, the marginal shifts observed are consistent with the theoretical expectation that the hottest region of the planet should display the most water dissociation\cite{parmentier_thermal_2018}. This illustrates the utility of the \texttt{Eigenspectra} method to isolate parts of the dayside with stronger spectral features, with the potential to strengthen chemical detection significances.
%The optical opacity detection is similar to the result found for the full planet, which also showed a detection of H$^{-}$/TiO/VO but at a lower significance of $3.8\sigma$\cite{Coulombe2023}. 
%Such similarities are expected, as the ``hotspot'' region is both bright and directly visible throughout the observation, so it dominates the planet's dayside emission. The increased detection significance of optical opacity sources in the ``hotspot'' compared to the full planet may be because the ``hotspot'' spectrum is not diluted by the ``ring'' region, which does not show significant H$^{-}$/TiO/VO opacity, whereas the full planet result incorporates dilution from the ``ring'' region. This shows the utility of the \texttt{Eigenspectra} method in increasing detection significance by separating parts of the dayside from which a specific feature originates.

%Attempted re-write based on moving most of this to the methods
%In contrast, similar full atmospheric inferences of the ring group show a non-inverted temperature structure, and t
The ring group spectrum is qualitatively similar to GCM predictions, which display water emission features but at a brightness temperature $\approx500$~K colder than the hotspot region (see Supplementary Figure \ifthenelse{\equal{\arxiv}{y}}{\ref{fig:eigen_spectra_check}}{2}). %clearly at odds with the full dayside retrieval and the GCM predictions. 
However, we also found that the retrieved T-P profile and chemical abundances for the ring group depended sensitively on how the models account for geometric effects such as %the exact model chosen and may be complicated by several factors such as 
the different average line of sight through the atmosphere in the ring group compared to a standard full-dayside secondary eclipse. We discuss the ring group results further in the Methods.

%This work presents the first three-dimensional map of an exoplanet atmosphere, providing an unprecedented level of detail in the spatial temperature structure and compositional changes. The large wavelength coverage and high precision of JWST make similar 3D mapping studies possible for a wider population of exoplanet atmospheres. These 3D maps enable detailed comparisons with theoretical predictions from GCMs, placing crucial constraints on our understanding of the underlying physics driving the maps we observe.

%{\color{red}FINDME revisit at end - Megan edited the last couple sentences}
%This work presents the first observational evidence for the three-dimensional nature of exoplanet atmospheres, identifying the horizontal and vertical temperature gradients of an ultra-hot Jupiter. 
While previous observations with Spitzer, Hubble, and JWST led to key advancements in our understanding of hot exoplanets, they were fundamentally limited to hemisphere-integrated spectra\cite{Knutson2014,Stevenson2014, Stevenson2017,MikalEvans2023,ParmentierEtal2021mnrasCloudyPhaseCurves} or single wavelength photometric maps\cite{Majeau2012, DeWit2012, Rauscher2018,Coulombe2023}. %two-dimensional information (longitude + altitude or longitude + latitude). 
The spectroscopic eclipse mapping presented here is the first observational analysis to resolve multidimensional information at multiple wavelengths simultaneously. Our findings are consistent with predictions of water dissociation in the hottest part of the atmosphere\cite{parmentier_thermal_2018} and indicate the importance of hydrogen dissociation\cite{Tan2019} and/or nightside clouds\cite{Roman2019,Parmentier2021} in shaping substellar-to-limb temperature gradients. %day-night energy transport. 
Moving forward, the large wavelength coverage and high precision of JWST will enable similar multidimensional mapping for a large sample of exoplanet atmospheres, allowing the study of horizontal and vertical thermal and chemical gradients across a population of giant exoplanets. %3D properties of a population of exoplanets covering a range of atmospheric temperatures and compositions. 
Through comparison with theoretical predictions from GCMs, these maps will place crucial constraints on atmospheric dynamics and chemical transitions.%molecular dissociation create 3D spatial variations in the temperatures and compositions of exoplanet atmospheres. 

%These results shows the potential of spectroscopic eclipse mapping for identifying three-dimensional structures in the dayside atmospheres of highly irradiated exoplanets.

%Through comparison with theoretical predictions from GCMs, these 3D maps will place crucial constraints on our understanding of the underlying physics driving the maps we observe.

\begin{figure}
    \centering
    \ifthenelse{\equal{\arxiv}{y}}{
    \includegraphics[width=\linewidth]{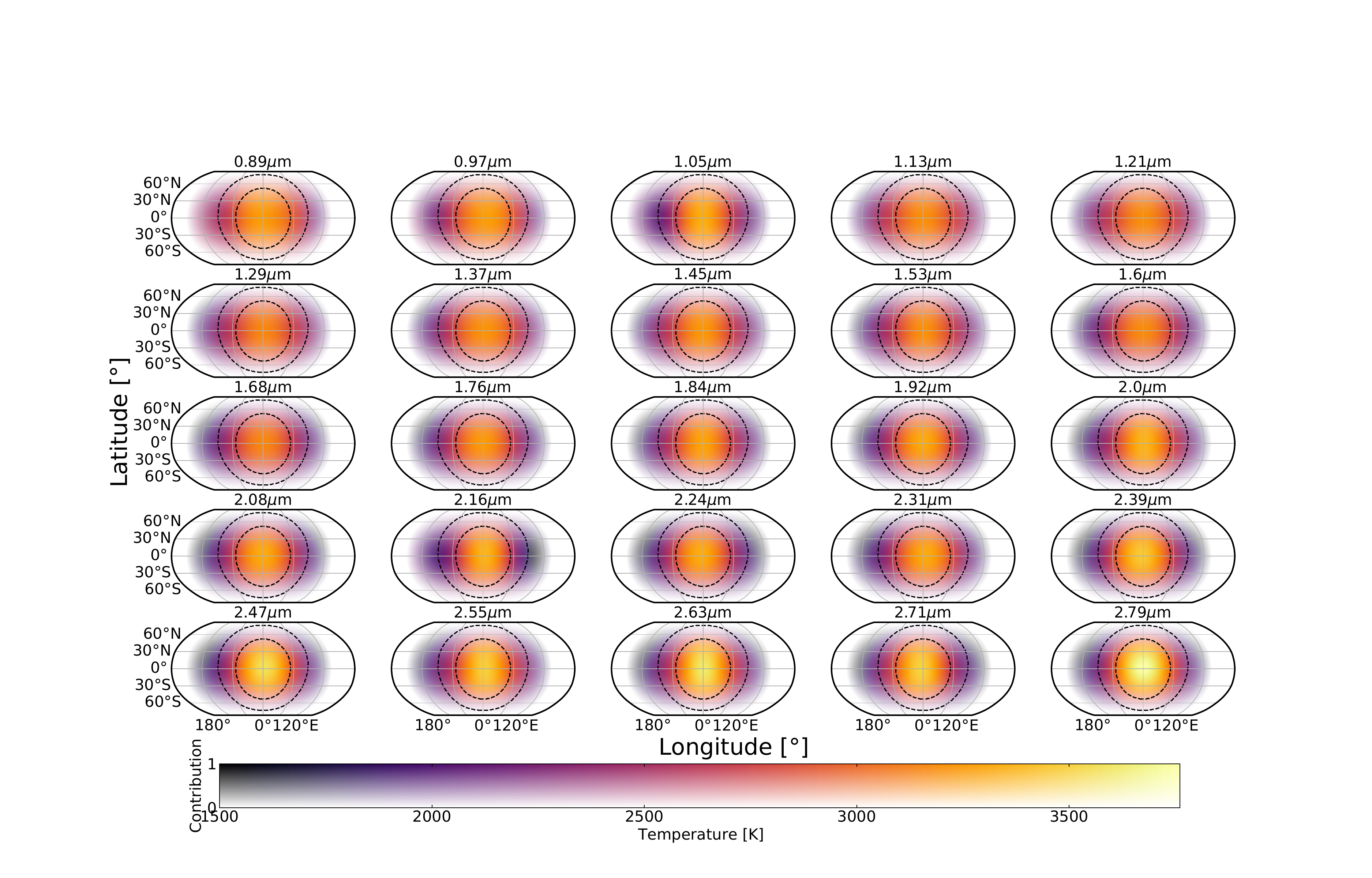}
    }{}
    \caption{\sep Two-dimensional maps %(left-hand panels) and light curve fits (corresponding right-hand panels) 
    from the \texttt{Eigenspectra} method for each of the 25 spectroscopic bins. %On the projected maps, 
    Colors indicate the temperature, while transparency indicates the relative contribution to the overall observed flux at the point of maximum visibility, based on the angle between a given point on the map and the line of sight to the observer. A maximum contribution of 1 indicates a latitude/longitude that is at the sub-observer point at some point during the observations. Dotted black curves delineate the three regions identified by the \texttt{Eigenspectra} mapping method. %On the light curves, black points with error bars indicate wavelength and time-binned, systematics-corrected data from Coulombe et al. (2023)\cite{Coulombe2023}, and red %solid and blue dashed 
    %lines show best fits from the \texttt{Eigenspectra} method. The light curves are shown in planet flux ($F_p$) divided by stellar flux ($F_s$). The models fit the data well and show differences in brightness with wavelength, showing 
    Evidence of multidimensional atmospheric structure can be seen in the varying hotspot temperature and shape with wavelength.} %in the data.} %and \texttt{ThERESA} methods, respectively. At many points the fits are similar enough that only one line can be seen by eye.}
    \label{fig:maps_8bin}
\end{figure}

\begin{figure}
    \centering
    \ifthenelse{\equal{\arxiv}{y}}{
    \includegraphics[width=\linewidth]{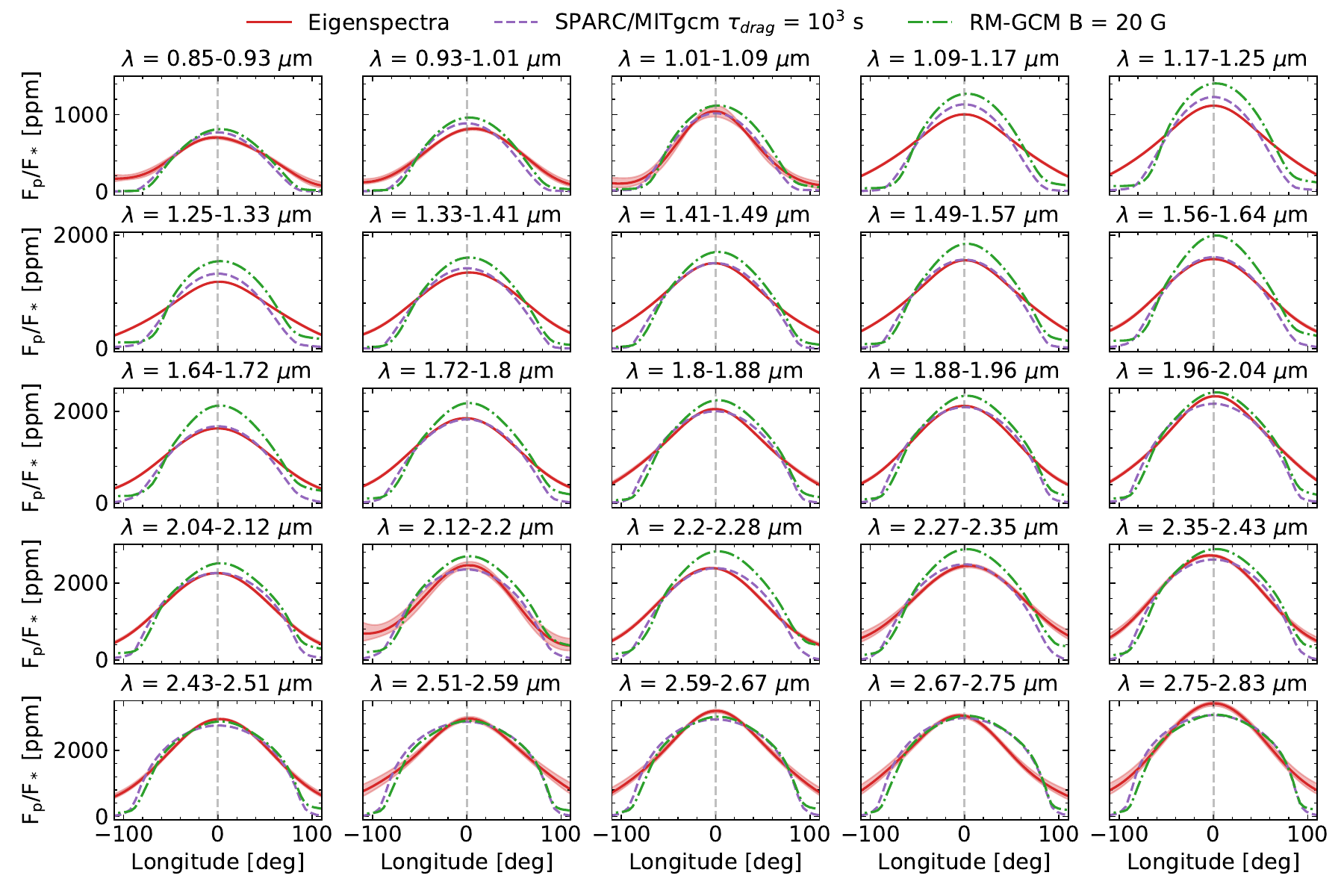}
    }{}
    \caption{\sep Retrieved longitudinal profiles at each wavelength range compared against GCMs. The \texttt{Eigenspectra}-retrieved profiles and GCMs including drag both show small hotspot offsets and sharp temperature gradients away from the substellar point at all wavelengths. The red lines and regions show the median retrieved longitudinal profiles and their 1$\sigma$ confidence intervals, respectively, measured with \texttt{Eigenspectra} for the twenty-five spectral bins considered. The profiles are obtained by weighting the retrieved 2D maps by the squared cosine of the latitude. The profiles are compared to two GCMs from ref.~\cite{Coulombe2023} that matched the white-light map well - the SPARC/MITgcm (purple dash-dot line), which has uniform drag of timescale $\tau_{\text{drag}}$ = 10$^3$~s, and the RM-GCM (green dashed line), which includes a kinematic magnetohydrodynamical drag model with an internal magnetic field of $B \sim 20$~G. We note that the GCMs as shown here are processed to remove the ``null space'' of components which are physically inaccessible to eclipse mapping (see Methods and refs. \cite{Luger2021, ChallenerRauscher2023ajNullSpace}). Vertical dashed lines indicate zero longitude.}
    \label{fig:longprof}
\end{figure}

\begin{figure}
    \centering
    \ifthenelse{\equal{\arxiv}{y}}{
    \includegraphics[width=0.98\linewidth]{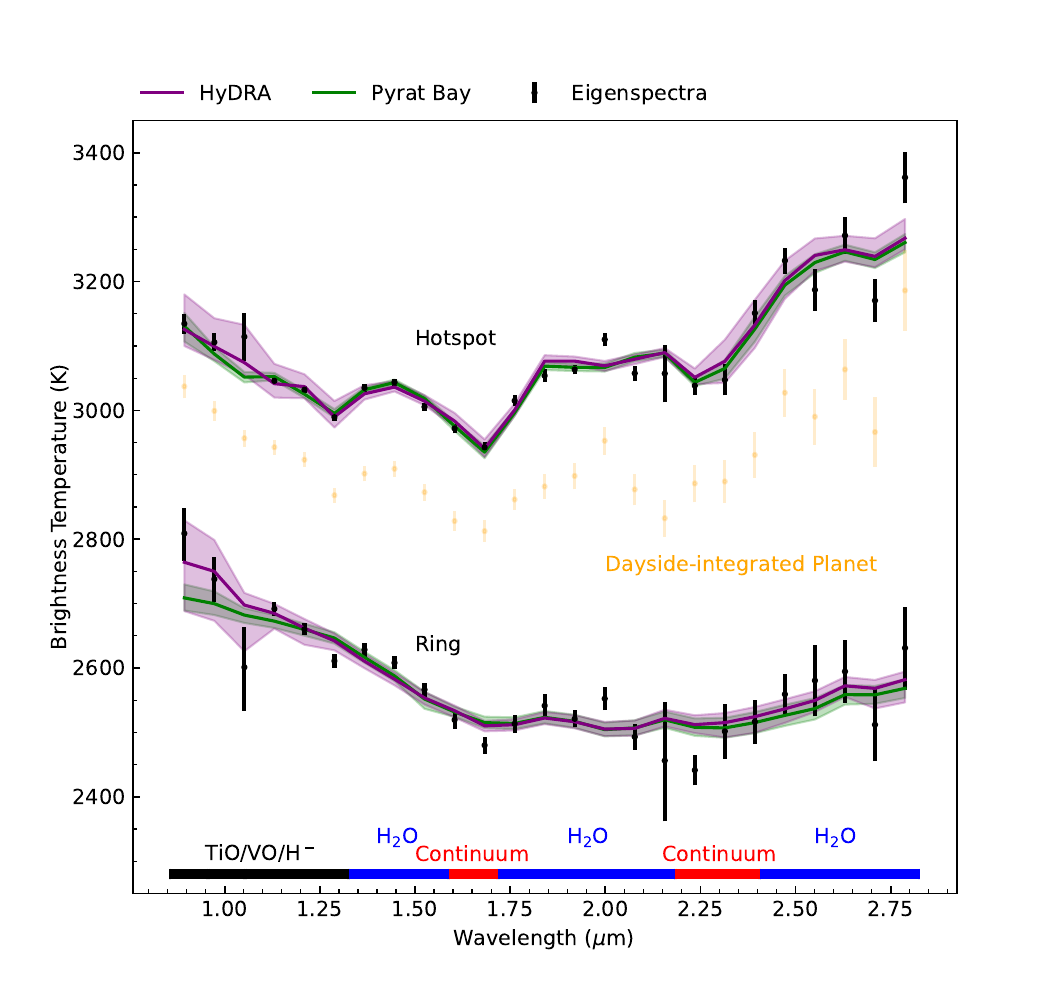}
    }{}
    \caption[]{\sep Hotspot and ring group spectra from \texttt{Eigenspectra} bracket the full dayside-integrated spectrum. Black points with error bars (standard deviation; see Methods) show the \texttt{Eigenspectra} emission spectra from the hotspot and ring groups, while yellow points show the hemispherically-averaged homogeneous dayside emission spectrum \cite{Coulombe2023}, binned in wavelength to match the group spectra. The spectra have been converted to brightness temperature by assuming blackbody emission for the planet in each bin and a PHOENIX emission model for the star. The \texttt{HyDRA} (purple) and {\pyratbay} (green) regions show 95.45\% (2$\sigma$) credible regions from 1D atmospheric retrievals on each spectrum. Labels along the bottom show the wavelength ranges at which different atmospheric constituents create features in the spectrum. %The ``hotspot'' and ``ring'' spectra bracket the dayside average, and 
    An average of the flux from the hotspot and ring regions produces a spectrum matching the dayside average (Extended Data Figure \ifthenelse{\equal{\arxiv}{y}}{\ref{fig:eigen_spectra_check}}{2}). See the Methods for further discussion of the ring spectrum and the mismatch with associated retrievals.}
    \label{fig:spectra}
\end{figure}

\clearpage

%######################################################################################################################

\begin{methods}

%{\color{red}FINDME MEGAN: Ryan's text about the difference from a uniform fit}

We applied two complementary spectroscopic eclipse mapping methods to the data: \texttt{Eigenspectra}\cite{Mansfield2020} and \texttt{ThERESA}\cite{ChallenerRauscher2022}. We used two methods because, as described below, they interpret the data in distinct ways, giving us a way to check which conclusions derived from the eclipse maps are robust to differences in mapping methods. 
While \texttt{Eigenspectra} fits the spectroscopic data well, \texttt{ThERESA} struggled to match the emission features in the data, as \texttt{ThERESA} simultaneously fits a 3D model to all the spectroscopic lightcurves, giving the model less flexibility than fitting each spectroscopic bin individually.
Therefore, we chose to highlight only the \texttt{Eigenspectra} method in the main text, but we include \texttt{ThERESA} here as an independent check to verify some of the main results from \texttt{Eigenspectra}. Below, we describe some potential paths for future research to investigate how to improve multi-wavelength eclipse mapping. Both methods start with the wavelength-resolved, systematics-corrected light curves presented in Coulombe et al. (2023)\cite{Coulombe2023}.

\subsection{Mapping with \texttt{Eigenspectra}} The \texttt{Eigenspectra} method\cite{Mansfield2020} splits 3D eclipse mapping into two stages. In the first stage, 2D brightness temperature maps are constructed for each wavelength bin following the eigenmapping method\cite{Rauscher2018}. First, we derive an orthogonal basis set of light curves from spherical-harmonic light curves using principal component analysis, where each light curve has a corresponding 2D map component. Then, we perform a 2D fit at each wavelength using a subset of this basis set of light curves. The fitting begins with a small number of components, and the number of components is increased until the Bayesian Information Criterion (BIC) indicates that the addition of more components is not preferred. For the fits presented in the main text, the 2D mapping preferred 4 or 5 free parameters at each wavelength, except for the fit at $1.05$~$\mu$m, which preferred 7 free parameters. To perform multi-wavelength mapping, \texttt{Eigenspectra} then extracts spectra from a grid of points in latitude and longitude across the visible area of the planet and uses a $k$-means clustering algorithm to identify regions of the planet with similarly-shaped spectra. This grouping is repeated in an MCMC framework to estimate uncertainties in the resulting grouped %per-grid-point 
spectra.

The number of distinct groups is chosen by starting with one group and increasing the number one by one until %and choosing 
the largest number of groups for which individual pixels are sorted into the same group across 75\% of the MCMC map iterations is identified. This ensures that the number of groups is limited by the ability of the data to precisely sort latitude/longitude points into the best-fit group. We note that the 75\% cut-off was chosen arbitrarily based on visually inspecting the results for different numbers of groups. Future work to perform \texttt{Eigenspectra} mapping on a larger sample of planets should further investigate whether this cut-off holds across different data sets. Supplementary Figure \ifthenelse{\equal{\arxiv}{y}}{\ref{fig:eigen_group_histos}}{1} shows the mean group maps and histograms of the assigned group for randomly chosen points on the map for the 25-bin analysis described in the main paper. The data showed a clear separation between groups when using 2 or 3 groups, but became mixed when using 4 groups.%, so we used 3 groups for the analysis.

We note that the grouping performed by \texttt{Eigenspectra} creates discrete spectra, but these discrete spectra likely represent a true planet with continuous properties, such as a smooth temperature gradient. The $k$-means clustering allows us to set the number of groups by the precision of the data, such that a more precise data set would be able to identify more spectra, because the change in properties required to distinguish two spectra with smaller error bars is correspondingly smaller. In the limit of infinitely precise data, each latitude/longitude point would be identified as a distinct group with distinct properties. However, this discrete representation allows us to determine how much the properties change across the visible area of the planet in a way that is 
 regulated by the signal-to-noise of the real observational data. 

For each identified group, the \texttt{Eigenspectra} method then creates a representative spectrum by taking an area- and visibility-weighted mean of the spectrum of each point included in the group, and scaling it by an area weighting to represent it on the same scale as a regular secondary eclipse spectrum, which covers a full visible hemisphere of a planet. The primary outputs of this mapping method are therefore a handful of spectra representing emission from different regions on the planet, which are run through atmospheric inference (retrieval) codes to measure molecular abundances and thermal structures. By virtue of the area and visibility weighting, the Eigenspectra are mathematically defined such that they should produce a hemisphere-integrated brightness equivalent to that of the full eclipse spectrum. Supplementary Figure \ifthenelse{\equal{\arxiv}{y}}{\ref{fig:eigen_spectra_check}}{2} shows that the hemisphere-integrated brightness indeed matches the wavelength-binned eclipse spectrum from ref.\cite{Coulombe2023}.%, which indicates that the \texttt{Eigenspectra} method is fitting the light curves well.

The error bars on the grouped spectra are calculated by taking the standard deviation of each point included in a group across all MCMC realizations of the planet map. The MCMC runs used 100 walkers, 7000 steps, and a burn-in of 700 steps. Convergence was evaluated by ensuring that the chain was at least 50 times as long as the autocorrelation timescale (see the \texttt{emcee}\cite{Foreman-Mackey2013} documentation at https://emcee.readthedocs.io/en/stable/tutorials/autocorr/). After calculating the errors in this way, we found that the hemisphere-integrated brightness had slightly smaller error bars than those from the original dayside eclipse spectrum from ref.\cite{Coulombe2023}. We tested running retrievals in the same format as the fiducial retrievals but with the error bars scaled up by a factor to match the original dayside eclipse spectrum, and we found that this change did not impact the retrieval results.

The method described above closely follows the method for mapping with \texttt{Eigenspectra} described in ref.\cite{Mansfield2020}, with two key improvements. First, for the 2D mapping with the eigencurves method, we restricted allowed planet maps to those which produce positive fluxes at all observed latitudes and longitudes, as a realistic planet must have positive thermal emission. Second, the area-weighting was applied to the resulting mean spectra for each group in order to allow atmospheric retrieval with standard secondary eclipse retrieval codes. We computed per-point spectra on a grid with a resolution of 1$\degree$ in both latitude and longitude. We also tested grids with a resolution of 3$\degree$ and 9$\degree$ and found that the positions of the groupings did not depend on the grid resolution.

The analysis described in the main text used 25 wavelength bins evenly spaced between $0.85-2.83$~$\mu$m, with a width of $0.079$~$\mu$m. We achieve reduced $\chi^2$ values of $1.02-1.39$ for the single-wavelength eigenmapping fits, with between 4 and 7 free parameters per fit and 2719 data points, and an overall $\chi^{2}_{\nu}=1.19$ for the full multi-wavelength eigenmapping fit. The best fits at each wavelength were obtained with a small number of eigenmapping components, restricting the resulting maps to the large-scale patterns characteristic of low-order spherical harmonics. The $\chi^2_{\nu}$ values at each wavelength are slightly above the expected value of 1 for a fit with correctly-estimated error bars. These elevated values are likely because the spectroscopic light curves were corrected for systematics at a higher resolution by ref.\cite{Coulombe2023} and then later binned down for use in this work. We recommend that future work investigate removing systematics at the same wavelength resolution at which eclipse mapping fits are performed, and/or performing simultaneous systematics and eclipse mapping fits.

In addition to the 25-wavelength bin fit described in the main text, we tested whether the results depended on the wavelength resolution by running a lower-resolution \texttt{Eigenspectra} fit. The lower-resolution fit had 8 wavelength bins, with their central wavelengths and widths optimally chosen to capture spectral features seen in the original secondary eclipse spectrum (see Supplementary Table \ifthenelse{\equal{\arxiv}{y}}{\ref{tab:eigenspectra}}{1}). 
The light curve fits for the 8 wavelength bins had reduced chi-squared between $1.22-2.26$, with between 4 and 6 free parameters per fit and 2719 data points. The larger reduced chi-squared values are likely due to the greater amount of binning applied to the original data. We found that the temperature maps had the same shape as for the 25-bin fit, and the \texttt{Eigenspectra} method still identified 3 distinct spectral groups in nested rings. Additionally, atmospheric retrievals on the 8-bin hotspot and ring groups showed consistent results with the 25-bin retrievals. 
We ultimately used the 25-bin spectrum for the main results because of the greater spectral resolution, but we used the 8-bin spectrum for comparison to the more computationally intensive \texttt{ThERESA} method, which could not be run on a greater number of wavelength bins within a reasonable timeframe. %larger wavelength bins provide tighter constraints on horizontal atmospheric temperature gradients while still allowing for vertical characterization.

%We also compared our mapping model with an unphysical model that treats the phase variation as a sine function and assumes the planet is spatially uniform, to attempt to disentangle the information content of the eclipse from the phase variation. \textbf{This model is unphysical because a truly uniform planet would have no phase variation.} 
%At many wavelengths, 
To provide a quantitative analysis of how much of our mapping information comes from the phase curve variation versus the eclipse itself, we compared our fit against one where we only allow for phase curve variation, represented by a double sine function, and assume a standard box-shaped eclipse with no additional perturbations to the shape of ingress or egress. We used a double sine function to match the fit to the out-of-eclipse variation performed by ref.\cite{Coulombe2023}. The double sine fit had 4 free parameters, comparable to the 4-7 free parameters in the \texttt{Eigenspectra} fits. This model is unphysical, since a planet with phase curve variation necessarily has spatial brightness gradients and so should also induce a signal during ingress and egress, but with this approach we are artificially requiring the eclipse shape to match that of a planet with uniform brightness. A comparison between these two fits then reveals how much signal is contributed solely from the eclipse. %\sout{At many wavelengths, 
%the difference between these models is small, indicating that the eclipse is largely consistent with a map that would be inferred from the phase variation alone, and the eclipse is serving to increase confidence in that measurement. The simple sinusoid fit achieves reduced chi squared values of $\chi^{2}_{\nu}=1.02 - 1.62$ in individual wavelength bins and $\chi^{2}_{\nu}=1.26$ for the full combined multi-wavelength fit.} 
We computed the BIC for both models and found a $\Delta \mathrm{BIC}$ between -3 and 713 depending on the wavelength, with a positive number indicating a preference for the \texttt{Eigenspectra} fit. For 17 of the 25 wavelength bins, the $\Delta \mathrm{BIC}>10$, indicating a strong preference for the \texttt{Eigenspectra} fit over the sinusoid fit. %At all wavelengths, the \texttt{Eigenspectra} fit produces an equal or lower reduced chi squared than a simple sinusoid fit. 
However, at some wavelengths, the improvement is marginal.
This lack of strong preference for the eclipse mapping fit over a sinusoidal fit is due to several factors, including: 1) WASP-18b rotates significantly during our eclipse observation, creating significant phase-curve variation that is present in a large part of the dataset relative to ingress and egress, %2) as an ultra-hot Jupiter with significant atmospheric drag, WASP-18b's dayside is longitudinally symmetric around a hotspot offset near 0$^\circ$, 
and 2) WASP-18b's low impact parameter reduces the strength of signatures of latitudinal temperature variation. However, the \texttt{Eigenspectra} analysis still provides multidimensional information that is not obtained from a simple sinusoidal phase variation fit - namely, the \texttt{Eigenspectra} fit reveals the radial extent of the hotspot so that its composition can be inferred separately from the surrounding dayside.
Other planets will likely be even better targets for 2D and 3D characterization with JWST\citep{BooneEtal2024MappingTargets}.

The component of eclipse mapping that can be uniquely inferred through secondary eclipses and not out-of-eclipse phase curve variation is latitudinal structure. Although the \texttt{Eigenspectra} maps show a lack of any latitudinal offset, this does not reflect an inability to constrain latitudinal information. To test the ability of the \texttt{Eigenspectra} fits to constrain latitudinal information, we follow methods similar to \cite{Lally2025} and artificially inject a latitudinal offset into the observations. Supplementary Figures \ifthenelse{\equal{\arxiv}{y}}{\ref{fig:ingresszoom}}{2} and \ifthenelse{\equal{\arxiv}{y}}{\ref{fig:egresszoom}}{3} show the light curves resulting from the minimum and maximum latitudinal offsets that produce fits with chi squared $\chi^{2} \leq10$ higher than the best fit in each wavelength bin. We found that this requirement results in a median constraint on the latitudinal offset of $-29\degree$ to $61\degree$. This comparison demonstrates that the \texttt{Eigenspectra} method would be able to detect latitudinal structure outside this range if such structure existed.

Supplementary Table \ref{tab:eigenspec25} lists the grouped 25-bin spectra resulting from the \texttt{Eigenspectra} analysis used in the main text, and Supplementary Table \ifthenelse{\equal{\arxiv}{y}}{\ref{tab:eigenspectra}}{1} lists the 8-bin spectra used for comparison to \texttt{ThERESA}. The hotspot, ring, and outer groups had a mean signal-to-noise of 483, 226, and 90, respectively. As described in the main text, although \texttt{Eigenspectra} identified three groups, we chose to fully analyze only two, the hotspot and ring. We chose not to apply atmospheric retrievals to the outer group because it had a signal-to-noise factor of $\approx2.5-12$ smaller than the other spectra and a much smaller contribution to the secondary eclipse signal. %We note that its signal-to-noise as reported in {\color{red}TABLE} 
While this may not appear to be a significant difference in signal-to-noise, it likely indicates that the shape of the outer group is being driven by the fitting method rather than the data. %but this is because of the application of the 2D eigenmapping method. 
%The 2D eigenmapping method applied at each wavelength essentially selects spherical-harmonic-based components to fit to a planetary map. 
As described above, the best 2D fit at each wavelength has a small number of eigenmapping components, and therefore will show only large-scale patterns characteristic of low-order spherical harmonics. Therefore, we suspect the shape of the outer group is primarily driven by the requirement of a smoothly varying map consistent with the much higher signal-to-noise hotspot and ring regions.%In many cases, the signal-to-noise may appear small because the fit confidently identifies a best number of eigenmapping components %Second, the outer group's spectral shape may have been influenced by the mapping method applied. The 2D eigenmapping constructs spherical-harmonic-based maps, so they are constrained to the relatively smooth shapes produced by lower-order spherical harmonics. This requirement, combined with the boundary condition that we require positive fluxes on the un-observed portion of the planet's nightside, likely influences the shape of the outer group, particularly because the shape of this group's spectrum is not as strongly constrained

%{\color{red}change this paragraph depending on what Figure 2 looks like}We note that the $0.96-1.06$-$\mu$m bin has larger uncertainties in the longitudinal profile than the other wavelength bins, as shown in Figure~\ref{fig:longprof}. This is because the 2D eigenmapping fit to this bin preferred a larger number of free parameters (4 versus 2 at most other wavelengths). This wavelength also had a longitudinal profile in closer agreement with the GCM, which could be physical or could be due to the greater flexibility provided by a larger number of free parameters. Future observations at higher signal-to-noise could allow fits with larger numbers of free parameters at other wavelengths as well.

%{\color{red}suggested supplementary figures: maybe 2D maps for 25 wavelength bins, supplementary videos scanning through maps at different wavelengths, actual light curves with fits overplotted (perhaps zoom-in on ingress and egress)}

\subsection{Mapping with \texttt{ThERESA}} Similar to \texttt{Eigenspectra},
\texttt{ThERESA} splits 3D eclipse mapping into two stages: 2D mapping and 3D mapping.
First, it constructs 2D star-normalized flux maps of the planet at each wavelength bin in the observation, using the eigenmapping method \cite{Rauscher2018}. This methodology is identical to 2D mapping with \texttt{Eigenspectra} (Supplementary Figure \ifthenelse{\equal{\arxiv}{y}}{\ref{fig:theresa2d}}{4}), although with ThERESA we use only 8 spectroscopic bins (the aforementioned lower-resolution fit, Supplementary Table \ifthenelse{\equal{\arxiv}{y}}{\ref{tab:eigenspectra}}{1}) to reduce the model complexity (see below).
To ensure physically-plausible maps, we enforce a positive-flux constraint at the longitudes that are visible during the observation (-134.7 -- 151.8 degrees).
To convert these flux maps into brightness temperature maps, we first compute a grid of planet brightness temperature vs.\ star-normalized planet flux, assuming the planet emits as a blackbody and using a PHOENIX\cite{Husser2013} model for the stellar spectrum.
Then we interpolate this grid to the fluxes in our observed maps to determine the brightness temperatures of the the maps.
Because these maps cover a relatively small wavelength range and our wavelength bins are chosen to probe small pressure ranges, converting these flux maps to brightness temperature maps is a reasonable choice for the 3D mapping (described below).

For the 3D mapping, \texttt{ThERESA} parameterizes the vertical placement of each of the 2D brightness temperature maps.
We test both a simple parameterization, where each 2D temperature map is placed at a single pressure level, and a more complex parameterization where the depth of the temperature map has a sinusoidal dependence on latitude and longitude, and the phase of the longitudinal sinusoid is allowed to vary.
Effectively, this sinusoidal model allows the photosphere to shift vertically with changing instellation and the resultant impact on temperature.
The 3D model also includes an internal temperature parameter that sets the temperature of the bottom of the atmosphere at all latitudes and longitudes.
We linearly interpolate, in log(pressures), along each column of the atmosphere between the 2D temperature maps and the internal temperature to create a 3D temperature grid.
The atmosphere is assumed to be isothermal above the highest-altitude 2D temperature map.

We then apply solar-abundance thermochemical equilibrium to each cell of the 3D temperature grid to calculate the atmosphere's chemical composition. 
For computational speed, we precompute a grid of chemical abundances vs.\ temperature and pressure using GGChem\cite{WoitkeEtal2018aandaGGchem}, and then interpolate to the temperatures in the model atmosphere.
Based in part on 1D atmospheric characterization of WASP-18b\cite{Coulombe2023}, we include H$_2$O, CO, CO$_2$, TiO, VO, and H$^-$ in the atmosphere.
We then calculate an emission spectrum from each column of the atmosphere using TauREx\cite{Al-RefaieEtal2019arxivTauRExIII} and integrate over the visible part of the atmosphere at each observation time, including the effects of planetary rotation, the angle between the sub-observer point and each grid cell, the area of each grid cell, and the occultation by the star.
We use ExoTransmit\cite{KemptonEtal2017paspExoTransmit} molecular opacities and compute the model at the opacity native resolution ($R \approx 1,000$), which is then binned to the data resolution. 
The model has 100 pressure layers evenly placed in log space between 0.0001 and 100 bar.

The resulting spectroscopic light curves are then compared against the data.
This process (3D temperature grid parameterization, composition calculation, emission spectra calculation, and spatial integration) is repeated behind an MCMC routine to explore parameter space.
For MCMC we use the MC3 package\cite{CubillosEtal2017apjRednoise}, which implements differential evolution Markov chains that can efficiently sample high-dimensional ($>50$) parameter spaces using a low number of chains\cite{terBraak2008}.
We use 7 chains and run a total of $\sim1.4$ million total iterations.
For comparison to the Eigenspectra mapping, we achieve autocorrelation lengths of $21 - 50$ for each parameter in the model.
We calculate contribution functions for each spectroscopic bin and apply a penalty to the model goodness-of-fit if the vertical positions of the 2D brightness temperature maps are inconsistent with the contribution functions\cite{ChallenerRauscher2022}.
This penalty is a confidence-region-like calculation where, if the vertical position of a given 2D map falls within the pressures where 68.3\% ($1\sigma$) of planetary emission at that wavelength originates, then there is effectively no penalty, but at significantly higher or lower pressures the penalty effectively causes the model to be rejected. 

The full 3D temperature map is shown in Supplementary Figure \ifthenelse{\equal{\arxiv}{y}}{\ref{fig:theresa3dtp}}{5}.
Broadly, the 3D temperature structures agree with \texttt{Eigenspectra}, with a thermal inversion near the substellar point that transitions to roughly isothermal near the limbs.
We achieve a reduced $\chi^2$ of 1.56 over all the spectroscopic light curves (33 model parameters, 21,752 data points), slightly worse than the \texttt{Eigenspectra} (Figure \ifthenelse{\equal{\arxiv}{y}}{\ref{fig:maps_8bin}}{1}) or the 2D \texttt{ThERESA} (Supplementary Figure \ifthenelse{\equal{\arxiv}{y}}{\ref{fig:theresa2d}}{4}) fits.
When considering all wavelengths together, the model residuals are well-behaved and distributed Gaussian-like around zero.
However, we note that \texttt{ThERESA} systematically overestimates or underestimates the light curves at certain wavelengths, and if the 3D model is post-processed into emission spectra (assuming thermochemical equilibrium and solar atomic abundances) from the planetary regions defined by \texttt{Eigenspectra}, the \texttt{ThERESA} spectra struggle to match the H$_2$O features in the data.
Models which create these emission features are within the parameter space we explored, but these models are rejected because they require increasing the temperature of the upper atmosphere, which leads to an overestimation of the total planetary emission, and such models also violate the contribution-function-consistency criterion.
Because of this discrepancy with the observed spectra, we opt to only report the \texttt{Eigenspectra} results in the main text, which do not experience similar difficulties in fitting the spectroscopic data.

This mismatch motivates several avenues for additional work to understand 3D atmospheric retrieval with JWST data.
First, \texttt{ThERESA} assumes that the planet's upper atmosphere is isothermal, an assumption that worked well for synthetic data based on GCMs\cite{ChallenerRauscher2022} but may depress molecular emission features necessary to fit these data.
Adjustments to the thermal profile parameterization that allow for flexibility in upper atmospheric temperature gradients would likely give the model the capability to match stronger molecular features. 
Second, \texttt{ThERESA} aims to place 2D brightness temperature maps at the pressures corresponding to contribution function maxima, to prevent non-physical scenarios where the 2D maps are placed at extremely high or low pressures. 
In reality, these 2D brightness temperatures come from a range of pressures, so the corresponding emission over that range, not just the emission at the peak of the contribution function should be consistent with the 2D maps.
Modifying the contribution function consistency check in this way would likely also reduce the model's chances of creating an extended isothermal upper atmosphere.
Finally, for simplicity, \texttt{ThERESA} assumes thermochemical equilibrium at solar atomic abundances. 
Expanding this framework to fit for bulk metallicity and C/O ratio, for example, could give the model some of the additional flexibility it needs to fit the data.

\subsection{Eclipse-mapping Null Space}

Some finer-resolution spatial flux patterns are inaccessible to eclipse-mapping analyses, as they create zero signal during the observation \citep{Luger2021, ChallenerRauscher2023ajNullSpace}. 
These patterns, collectively referred to as the eclipse-mapping null space, need to be removed from GCMs before comparing them against measured eclipse maps, as the measured maps will never place constraints on the null-space patterns.
The GCMs presented in Figure \ifthenelse{\equal{\arxiv}{y}}{\ref{fig:longprof}}{2} have been processed to remove the null space by representing the GCMs as high-degree spherical-harmonic maps, using principal component analysis to identify null components of the map, and removing those null components \citep{ChallenerRauscher2023ajNullSpace}.

%{\color{red}paragraph on future improvements to ThERESA}

\subsection{Retrievals on Eigenspectra}

\subsubsection*{\texttt{HyDRA}}
\label{sec:hydra}

\texttt{HyDRA} \cite{Gandhi2018,Gandhi2020,Piette2020,Piette2022} is an atmospheric retrieval framework which combines a parametric forward atmospheric model with a Nested Sampling Bayesian parameter estimation algorithm, \textsc{PyMultiNest} \cite{BuchnerEtal2014aaBayesianXrayAGN,FerozEtal2009mnrasMultiNest,Skilling2006}. The inputs to the forward model include six parameters for the temperature-pressure profile, and the deep-atmosphere abundances of each of the chemical species considered. In particular, we use the temperature-pressure profile parametrisation of ref.\cite{MadhusudhanSeager2009apjRetrieval}, which has been used extensively for atmospheric retrievals of exoplanet atmospheres, including ultra-hot Jupiters such as WASP-18b \citep{Gandhi2020,Coulombe2023}. The model also includes the abundances of chemical species which have opacity in the 0.8-2.8~$\mu$m range and are expected in H$_2$-rich atmospheres \cite{Madhusudhan2012,Gandhi2020}: collision-induced absorption (CIA) due to H$_2$-H$_2$ and H$_2$-He \citep{Richard2012}, H$_2$O \citep{Rothman2010}, CO \citep{Rothman2010}, CO$_2$ \citep{Rothman2010}, HCN \citep{Harris2006}, OH \citep{Rothman2010}, TiO \citep{McKemmish2019}, VO \citep{McKemmish2016}, FeH \citep{Dulick2003}, Na \citep{Burrows2003}, K \citep{Burrows2003} and H$^-$\citep{Bell1987,John1988}. For each opacity source, line-by-line absorption cross-sections\cite{Gandhi2017} are calculated using data from the references listed. The opacity from H$^-$ free-free and bound-free transitions is calculated using the methods of ref.\cite{Bell1987} and ref.\cite{John1988}, respectively. We additionally include the effects of thermal dissociation for H$_2$O, TiO, VO and H$^-$. The depletion in the abundances of these species is calculated as a function of pressure and temperature, using a parametric method\citep{parmentier_thermal_2018}. For all other species, the abundances are assumed to be constant with depth. In some of the retrievals, we additionally test the effects of adding a `dilution' parameter (an area covering fraction), which multiplies the overall emission spectrum by a constant factor between 0 and 1 \cite{TaylorEtal2020mnrasEmissionBiases}. 

The forward model computes the thermal emission spectrum of the atmosphere given the input parameters described above. The pressure range considered is 10$^{-5}$-10$^3$~bar. The spectrum is calculated at a resolving power of $R \sim 15\,000$, and is convolved to the resolution of the instrument before being binned to the data resolution. The binned model is compared to the data to calculate the likelihood of the model instance. We use 2000 live points in the Nested Sampling parameter estimation algorithm. \texttt{HyDRA} ultimately outputs the posterior probability distributions for each model parameter, from which we calculate the median and 1-sigma contours for the retrieved spectrum, temperature profile and chemical abundance profiles. We additionally perform Bayesian model comparisons to determine the evidence for one model (e.g., including a particular molecule) over another (e.g., which excludes that molecule). To do this, we compare the Bayesian evidence from the retrievals using each model, which we convert to a `sigma' confidence value using the methods of ref. \cite{Benneke&Seager_2013}.

\subsubsection*{{\pyratbay} modeling framework}
% Patricio / Jasmina

{\pyratbay} is an open-source framework that enables atmospheric
modeling, spectral synthesis, and Bayesian retrievals of
exoplanet observations\cite{CubillosBlecic2021mnrasPyratBay}.
The atmospheric model consists of 1D parametric profiles of the
temperature, volume mixing ratios (VMR), and altitude as a function of
pressure (hydrostatic equilibrium).  For this analysis, we considered a pressure array extending
from $10^{-9}$ to 100~bar and a wavelength array from 0.8 to 3.0
{$\mu$m} sampled at a resolving power of $R=15\,000$.  The temperature
profile follows the parametric prescription of
ref\cite{MadhusudhanSeager2009apjRetrieval}.  Our framework computes
abundances in thermochemical equilibrium via a Gibbs free-energy
optimization code that combines the flexibility and performance of
previous chemical frameworks\cite{BlecicEtal2016apsjTEA, AgundezEtal2012aaChemistry}.  This chemical code produces VMR profiles consistent with the pressure, temperature, and elemental composition of the atmosphere at each layer.  The chemical network
includes 45 neutral and ionic species that are the main carriers of H,
He, C, N, O, Na, Si, S, K, Ti, V, and Fe.  We adopted three free
parameters to vary the elemental composition at each iteration: a
carbon-abundance scaling factor ([C/H], relative to solar values), an
oxygen scaling factor ([O/H]), and a third ``catch-all'' parameter
that scales the abundance of all other metals ([M/H]).  The altitude
of each layer is calculated assuming hydrostatic equilibrium.
Finally, we also considered a free parameter for dilution\cite{TaylorEtal2020mnrasEmissionBiases}, which accounts for
spatial inhomogeneities of the planetary flux.

% Opacities
For a given set of atmospheric parameters, {\pyratbay} computes the emission spectrum
considering opacities from the Na and K resonant
lines\cite{BurrowsEtal2000apjBDspectra}; H, \ch{H2}, and He Rayleigh
scattering\cite{Kurucz1970saorsAtlas}; \ch{H2}--\ch{H2} and
\ch{H2}--He collision-induced
absorption\cite{BorysowEtal1988apjH2HeRT,
  BorysowFrommhold1989apjH2HeOvertones, BorysowEtal1989apjH2HeRVRT,
  BorysowEtal2001jqsrtH2H2highT, Borysow2002jqsrtH2H2lowT,
  JorgensenEtal2000aaCIAH2He}; \ch{H-} free-free and bound-free
opacity\cite{John1988aaHydrogenIonOpacity}; and molecular line lists
for CO, VO, \ch{H2O}, and TiO \cite{LiEtal2015apjsCOlineList,
  McKemmishEtal2016mnrasVOMYTexomolVO,
  PolyanskyEtal2018mnrasPOKAZATELexomolH2O,
  McKemmishEtal2019mnrasTOTOexomolTiO}.  To process the large
molecular line-list opacity files, we applied the \textsc{repack}
package\cite{Cubillos2017apjRepack} to extract the dominant line
transitions, which we then sampled over a temperature, pressure, and
wavelength grid for interpolation during retrieval runs.
% SAMPLER
The Bayesian sampling in {\pyratbay} is managed with the \textsc{mc3}
package\cite{CubillosEtal2017apjRednoise}, in this case using the
\textsc{MultiNest} nested sampling
algorithm\cite{FerozEtal2009mnrasMultiNest,
  BuchnerEtal2014aaBayesianXrayAGN} with 1500 live points.
% utilizing 2000 live points, respectively.  The inputs of the models
% are the known system parameters, a stellar spectrum
% \citep{Castelli&Kurucz2003IAUS}, and the atmospheric elemental
% composition.

\subsubsection*{Hotspot group retrievals}
%{\color{red}suggested supplementary figures: full posteriors for the groups 1 + 2 fits}

Supplementary Figures \ifthenelse{\equal{\arxiv}{y}}{\ref{fig:hotspot_pairs_pyratbay}}{6} and \ifthenelse{\equal{\arxiv}{y}}{\ref{fig:pairs_hydra}}{7} show the results from retrievals on the hotspot group. Both retrievals of the hotspot group find a strong thermal inversion
around the $\sim$1~bar pressure level, where the temperature increases
from 2900 to 3300~K.  Above this level most molecules start to
thermally dissociate, depleting the upper layers of the main
optical/NIR absorbers (\ch{H2O}, TiO, and \ch{H-}).  The retrieved
spectra %match the data well ({\color{red}$\chi^2=XX$}) and 
are dominated by a
series of \ch{H2O} emission bands at $\lambda>1.25$~{\microns} and by
optical opacity (e.g., H$^{-}$, TiO, and/or VO) at $\lambda<1.5$~{\microns}. %, which are
%probed between the 10--0.1~{bar} range.  We observe traces of TiO and
%CO absorption features (at 0.95 and 2.4~{\microns}, respectively),
%although they do not significantly impact the fit.  
%{\color{red}check these numbers for the 25 bin retrievals. (PC: updated!)} 
The {\pyratbay} retrieval shows a well-constrained posterior with sub-solar elemental
abundances (${\rm [M/H]}=-0.22\pm0.16$) and a sub-solar C/O ratio (${\rm C/O}=0.22 \pm 0.15$); these elemental compositions lead to a water abundance of $\log_{n_{\rm{H_{2}O}}}=-3.20\pm0.17$ at the photosphere. Similarly, the \texttt{HyDRA} retrieval shows a well-constrained water abundance of $\log_{n_{\rm{H_{2}O}}}=-3.7^{+0.3}_{-0.2}$, although it is unable to precisely constrain the abundances of any other species. These results generally agree with the full dayside atmospheric constraints\cite{Coulombe2023}, which is expected as the bright and directly visible hotspot dominates thermal emission throughout the observation.

\subsubsection*{Ring group retrievals}

%Nominal retrieval shows a non-inverted T-P profile, but no constrained abundances. Lack of constrained abundances means the T-P profile is likely not strongly constrained to be any specific shape, as we would not be sensitive to a wide range of atmospheric pressures.
%We have reasons to believe this retrieval might be missing something - by eye there look to be slight H2O emission features

Supplementary Figures \ifthenelse{\equal{\arxiv}{y}}{\ref{fig:retrieve_ring}}{8} and 
\ifthenelse{\equal{\arxiv}{y}}{\ref{fig:ring_spectra}}{9} show a summary of retrieved constraints for the ring group. We note that we saw the same results for the ring group when using 8- and 25-wavelength bins. The nominal atmospheric retrievals of the ring group, as well as the \texttt{ThERESA} fit, result in physical
properties in stark contrast to the hotspot group, though this depends strongly on the model assumptions, as described below.  The nominal models result in
non-thermally inverted temperature profiles with brightness
temperatures of $\sim$2500--2700~K, probed mainly at pressures of
1--10~bar by the observations. This decrease in temperature from $\sim$3000--3200~K of the hotspot (Figure \ifthenelse{\equal{\arxiv}{y}}{\ref{fig:spectra}}{3}) is roughly consistent with the GCMs with atmospheric drag, although the GCM temperatures in the ring region vary significantly with latitude/longitude and show thermal inversions.
%%%
%While at pressures of $\sim$1 bar these temperatures differ by only
%$\sim$250~K from the hotspot temperature profile, at pressures above
%the 0.1 bar level the temperature differences quickly increase to
%$\gtrsim 800$~K.

%{\color{red}again check numbers with new 25 bin retrievals}
Perhaps the most puzzling outcome of the ring group retrieval is
the atmospheric composition. With {\pyratbay}, we found that the abundance posterior
distribution was constrained to the C/O$>1$ region, leading to extremely
low \ch{H2O} abundances ($VMR < 10^{-6}$), such that there were no
visible \ch{H2O} absorption bands in the model. Similarly, the nominal \texttt{HyDRA} retrievals on the ring group found very low H$_2$O abundances (also $VMR < 10^{-6}$), while we would expect $VMR \approx 10^{-3.3}$ (see Supplementary Figure \ifthenelse{\equal{\arxiv}{y}}{\ref{fig:retrieve_ring}}{8}). In both sets of retrievals, the ring spectrum was
mainly dominated by absorption due to \ch{H2}-\ch{H2} and \ch{H2}-He CIA. %, with a minor
%contribution from \ch{H-} at $\lambda<1.5$~{\micron} and CO at
%$\lambda\sim$2.3~{\micron}.
%%
This represents a drop of over two orders of magnitude in \ch{H2O} abundance
from the hotspot to the ring group. Such a steep gradient in dayside composition seems physically unlikely, especially since H$_2$O is expected to be more abundant in the cooler ring region compared to the hotspot, where thermal dissociation depletes the H$_2$O abundance. Additionally, the ring group spectrum appears by eye to show slight H$_{2}$O emission features at the same wavelengths where emission features are seen in the hotspot and full dayside spectra (e.g., slight peaks at $\sim$1.4~$\mu$m and $\sim$1.9~$\mu$m, Supplementary Figure \ifthenelse{\equal{\arxiv}{y}}{\ref{fig:ring_spectra}}{9}). It is, therefore, possible that H$_2$O absorption is shaping the ring group spectrum, but is incorrectly identified in the retrievals. Finally, the lack of any detected opacity aside from H$_{2}$-H$_{2}$ and \ch{H2}-He CIA in the ring group throws into question the validity of the retrieved T-P profile, as the retrievals would not be sensitive to a wide range of pressures without any species that can change the atmospheric opacity over the wavelengths we investigated.

We suspect that there are physical or geometric effects that the 1D
models are not able to capture, hence preventing the retrievals from providing a sound physical interpretation. We found that the standard model was strongly preferred over a simple blackbody ($>14\sigma$). The fact that the spectrum shows significant deviations from a blackbody indicates that the results of the standard retrieval are not due to an inability to detect atmospheric features; the data show a clear preference for a model with features over a perfect blackbody.

%As a test, we ran an additional \texttt{HyDRA} retrieval fitting the standard model with the addition of a dilution parameter, whose spectrum is also shown in Extended Data Figure~\ref{fig:ring_spectra}.
As a test, we ran additional retrievals fitting the standard model with the addition of a dilution parameter. Supplementary Figure \ifthenelse{\equal{\arxiv}{y}}{\ref{fig:ring_spectra}}{9} shows the spectrum for the \texttt{HyDRA} code. %First, we tested fitting the spectrum with a simple blackbody. We found that the standard model was strongly preferred over a simple blackbody ($>16\sigma$). The fact that the spectrum shows significant deviations from a blackbody indicates that the results of the standard retrieval are not due to an inability to detect atmospheric features; the data show a clear preference for a model with features over a perfect blackbody.
%Second, we tested fitting the standard model with the addition of a dilution parameter. 

One possible explanation for these results is that what may be a sharper boundary in spectral features between the hotspot and ring groups is smeared out by the 2D eigencurves fitting. Eigencurves fitting is based on maps constructed from relatively low-order spherical harmonics, so it is fundamentally limited to producing maps with relatively smooth gradients\cite{Rauscher2018,Mansfield2020}. If the true planet showed a rapid change in spectral features at a sharp boundary, the eigencurve mapping may smear out this sharp boundary, producing a mix of spectral features in the resulting group spectra that might confuse standard retrievals. However, we note that the grouped spectra are very similar in shape and amplitude to spectra derived from similar regions of a GCM, perhaps indicating that the eigencurve fitting does not have an oversized impact on the resulting spectra. The specific extent to which eigencurve fitting impacts the spectra can be investigated in the future by applying the \texttt{Eigenspectra} mapping method to GCM outputs where the ground truth map is known. %We explored a couple of
%these scenarios, but future research should investigate further the cause of the deviation from expected retrieval results in the ring group.

%\subsection{Viewing-angle shenanigans}

We also explored whether the slant viewing angle between the
observer and the flux from the ring group biases the retrievals. For
this, we modified the 1D emission models to, instead of integrating the
planet intensity over the entire day-side hemisphere, integrating only
over a region delimited by $\cos(\psi) \in (0.6, 0.2)$, where $\psi$
is the angle between the line of sight and the intensity vector
over the day-side hemisphere.  This is the region where the ring-group
flux originates.
% appropriately re-scaling the total flux by the projected areas, in
% the same manner as the eigenspectra approach does to produce the
% spectra.
While we confirmed that the slant viewing angle has a
wavelength-dependent impact on the emission spectra, we found no
significant changes in the retrieved temperatures or abundances
between this approach and the nominal retrieval. However, we have not ruled out the possibility that some other effect due to the non-standard geometry may be impacting the retrievals. Temperature variations within the ring group region could also affect the retrieval results. Indeed, the dilution parameter is designed to account for thermal inhomogeneities due to a hotspot region \cite{TaylorEtal2020mnrasEmissionBiases}, and could be compensating for variations within the relatively large ring group region, although imperfectly.

We exclude the chemistry results from the ring group from the main text due to our suspicions that the retrievals may be impacted by some combination of the factors listed above. While a more detailed investigation of these possibilities is outside the scope of this work, future research should investigate this further to improve upon spectroscopic eclipse mapping methods. Applying the \texttt{Eigenspectra} method to GCM outputs would allow an investigation of the effects listed above and whether improving upon any of them can increase the fidelity of the retrievals.

%\subsection{Retrievals with a dilution factor}

%\subsubsection{discussion of the dilution parameter/ring group retrievals}
%a couple paragraphs about what is going on here

%Retrievals on the ring group spectrum result in a non-inverted temperature profile and extremely low and/or unconstrained abundances of all chemical species included in the model. The model essentially fits the spectrum with only CIA opacity, which would require an unphysical gradient in metallicity across the dayside of the planet, going from slightly sub-solar abundances to pure H/He from the hotspot group to the ring group. This may indicate that the retrieval models do not include the necessary physics and/or geometric considerations to fit the data.

%\subsection{General circulation model post-processing}

%We post-process the SPARC/mit general circulation model presented in ref. \cite{Coulombe2023} to allow for comparison with the outputs from the retrievals performed on the ring and hotspot \texttt{Eigenspectra} groups.

%{\color{red}Louis-Philippe to fill in}

\end{methods}

%################################################################################################################

\noindent\textbf{Data availability}\\
The data used in this work are publicly available in the Mikulski Archive for Space Telescopes ({\small \url{https://archive.stsci.edu/}}). The data which was used to create all of the figures in this manuscript are freely available on Zenodo. \cite{W18bSpecEclMapZenodo}

\noindent{\textbf{Code availability}}\\
The open-source spectroscopic eclipse mapping pipelines used throughout this work are available at {\small \url{https://github.com/meganmansfield/eigenspectra}} (\texttt{Eigenspectra}) and {\small \url{https://github.com/rychallener/ThERESA}} (\texttt{ThERESA}). The {\pyratbay} atmospheric-retrieval package is available at \url{https://github.com/pcubillos/pyratbay}.

\begin{addendum}
 \item 
% Acknowledgments
This work is based on observations made with the NASA/ESA/CSA JWST. The data were obtained from the Mikulski Archive for Space Telescopes at the Space Telescope Science Institute, which is operated by the Association of Universities for Research in Astronomy, Inc., under NASA contract NAS 5-03127 for JWST. These observations are associated with program JWST-ERS-01366. Support for program JWST-ERS-01366 was provided by NASA through a grant from the Space Telescope Science Institute. M.W.M. acknowledges support through the NASA Hubble Fellowship grant HST-HF2-51485.001-A awarded by the Space Telescope Science Institute, which is operated by AURA, Inc., for NASA, under contract NAS5-26555, and from the Heising-Simons Foundation through the 51 Pegasi b Fellowship Program. R.C.C acknowledges support by a grant from the Research Corporation for Science Advancement, through their Cottrell Scholar Award, and this research was supported in part through computational resources and services provided by Advanced Research Computing at the University of Michigan, Ann Arbor. P.E.C. is funded by the Austrian Science Fund (FWF) Erwin Schroedinger Fellowship, program J4595-N. L.M. acknowledges financial contribution from PRIN MUR 2022 project
2022J4H55R. J.B. acknowledges the support received from the NYUAD IT High Performance Computing resources, services, and staff expertise, thanks to which a portion of this analysis was carried out.

 \subsection{Author Contributions Statement}
 All authors contributed significantly to one or more of the following: planning and construction of the observing proposal, project management, data analysis, theoretical modeling, and manuscript preparation.
 Many were instrumental in the preparation of the first publication on this observation.
 The following lists some specific contributions.
 NMB, 
 JLB and KBS provided leadership and management of the broad JWST Transiting Exoplanet Early Releease Science program.
 JLB, BB, and EMRK coordinated the ``Bright Star'' working group, which was responsible for the analysis of the WASP-18b eclipse observation.
 JLB, KBS, DKS, NC, JMD, JH, VP, and ES made significant contributions to the design of the program and observing proposal.
 DKS, ES, KBS, PEC, HB, JB, EMRK, VP, and RCC contributed to pre-launch planning activities, such as the Data Challenges.
 RCC, MWM, PEC, AAAP, and HB contributed significantly to writing the manuscript. 
 LPC provided the data products used in this analysis.
 RCC and MWM performed the \texttt{ThERESA} and \texttt{Eigenspectra} analyses, respectively.
 PEC, AAAP, and JB performed atmospheric retrieval analyses.
 RCC, MWM, LPC, PEC, and AAAP produced the figures in the manuscript.
 ER, HB, VP, and XT provided GCMs and/or aided with their interpretation.
 SLC, NI, LM, MCN, MR, MES, and LW provided significant feedback on the manuscript.
 
 \item[Competing Interests Statement] The authors declare that they have no competing interests.
 \item[Correspondence] Correspondence and requests for materials should be addressed to Megan Weiner Mansfield (email: mwm@umd.edu) or Ryan C. Challener (email:rcc276@cornell.edu).
\end{addendum}

%\newpage
%\subsection{Tables}

%\clearpage
%\subsection{Figure Legends/Captions}
%\clearpage

% If this is for arXiv, then we want the extended data figures in this file. If it's for submission, they want it in a separate file. I've just used placeholder empty figures for submission so that the labels come out right when the file compiles.

% If this is for arXiv, then we want the supplementary information in this file. If it's for submission, they want it in a separate file (supplementary-nature.tex).

%###FIGURES############################################################################################################

%---Supplementary Information-----------------------------------------------------------
\clearpage
\setcounter{page}{1}
\setcounter{figure}{0}
\setcounter{table}{0}
\renewcommand{\figurename}{Supplementary Fig.}
\renewcommand{\tablename}{Supplementary Table}

\begin{center}
\textbf{\huge{}Supplementary Figures}{\Huge\par}
%\textbf{\huge{}Extended Data}{\Huge\par}
\par\end{center}

\begin{figure}[h]
    \centering
    \includegraphics[width=\linewidth]{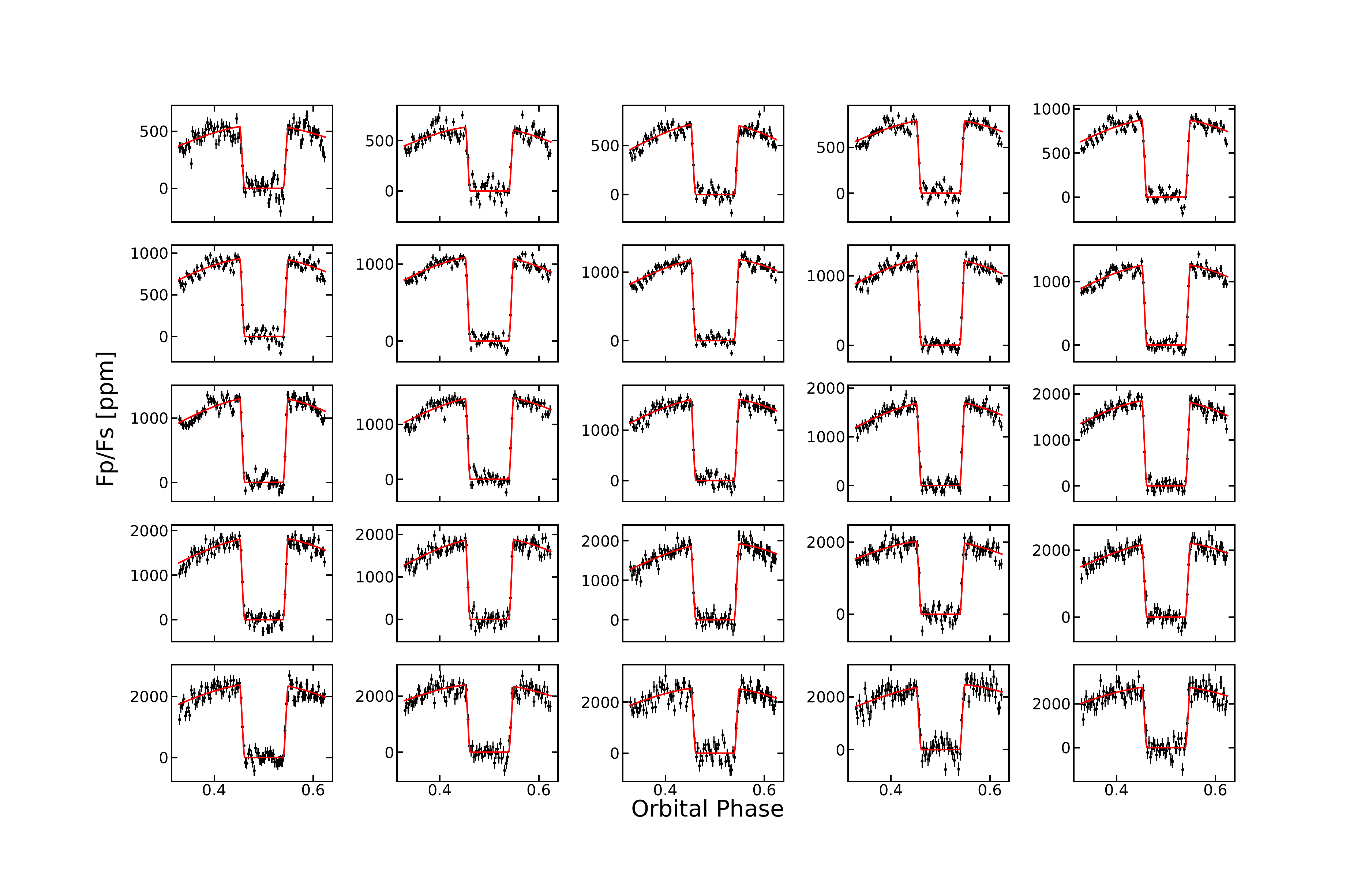}
    \caption{\sep Light curve fits from the \texttt{Eigenspectra} method for each of the 25 spectroscopic bins. Black points with error bars indicate wavelength and time-binned, systematics-corrected data from Coulombe et al. (2023)\cite{Coulombe2023}, and red %solid and blue dashed 
    lines show best fits from the \texttt{Eigenspectra} method. The light curves are shown in planet flux ($F_p$) divided by stellar flux ($F_s$). The models fit the data well and show differences in brightness with wavelength, showing evidence of multidimensional atmospheric structure in the data.} %and \texttt{ThERESA} methods, respectively. At many points the fits are similar enough that only one line can be seen by eye.}
    \label{fig:lightcurve25}
\end{figure}

\vspace{3cm}
\begin{figure}[h]
    \centering
    \includegraphics[width=0.35\linewidth]{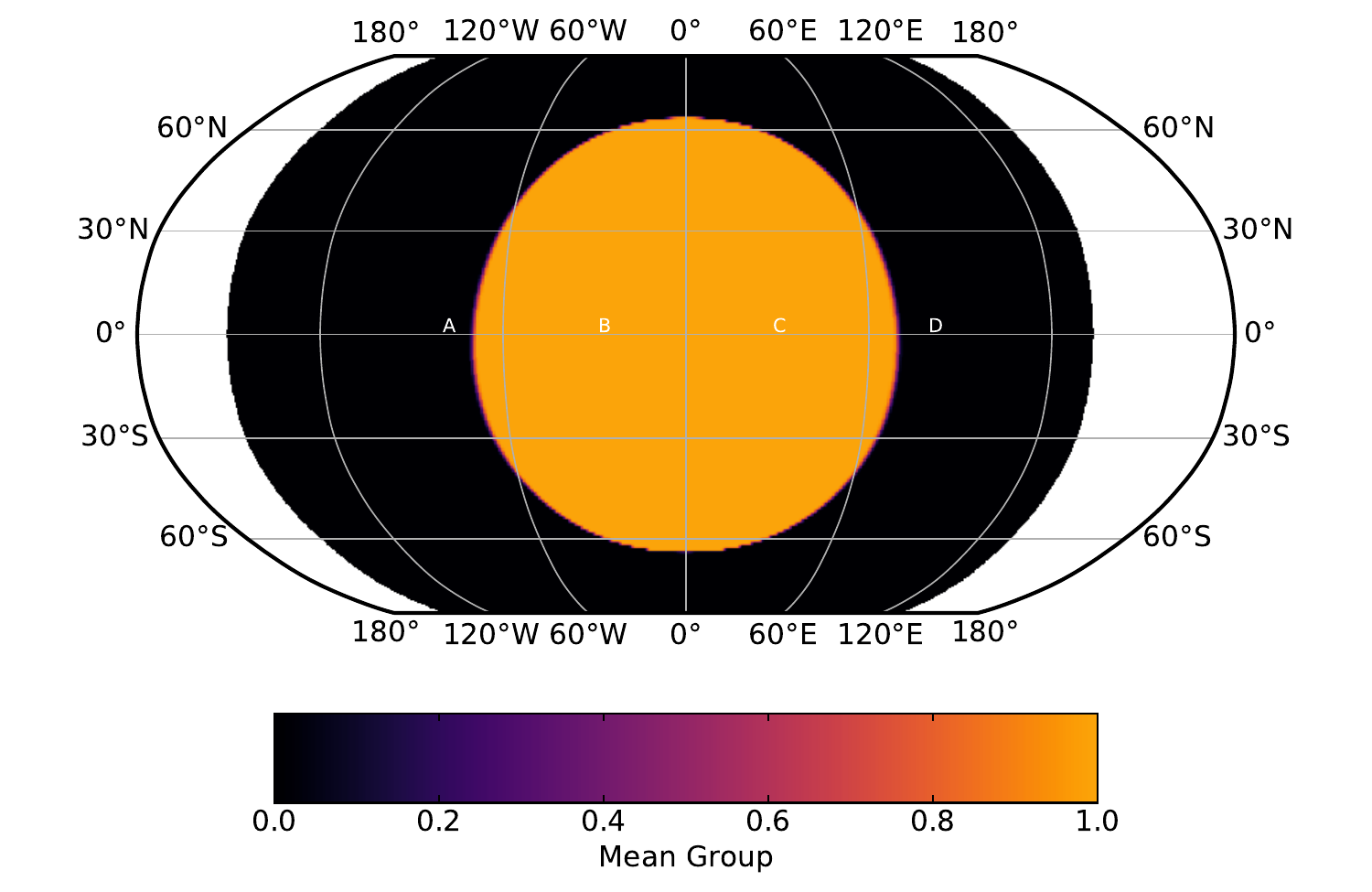}
    \includegraphics[width=0.35\linewidth]{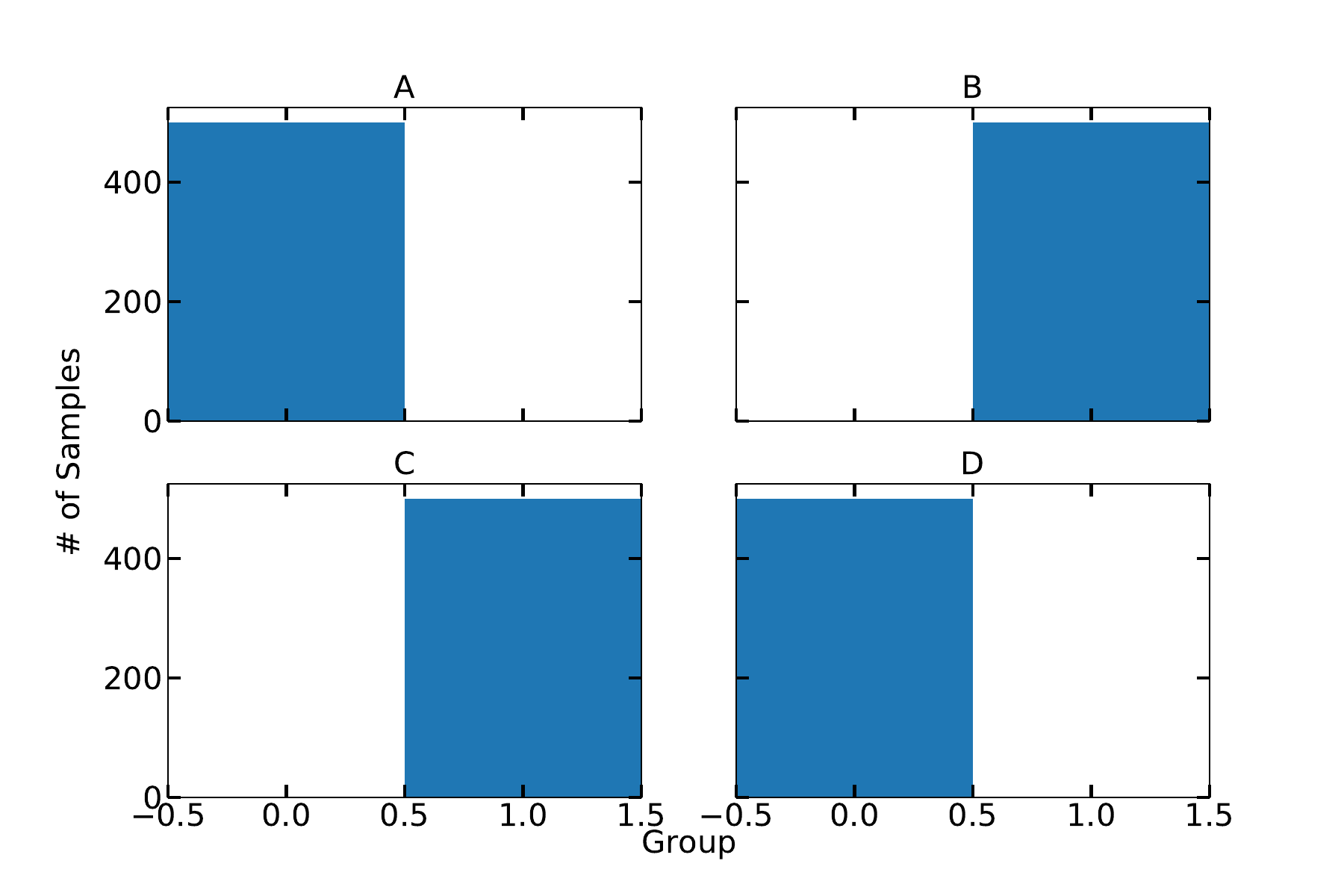}
    \includegraphics[width=0.28\linewidth]{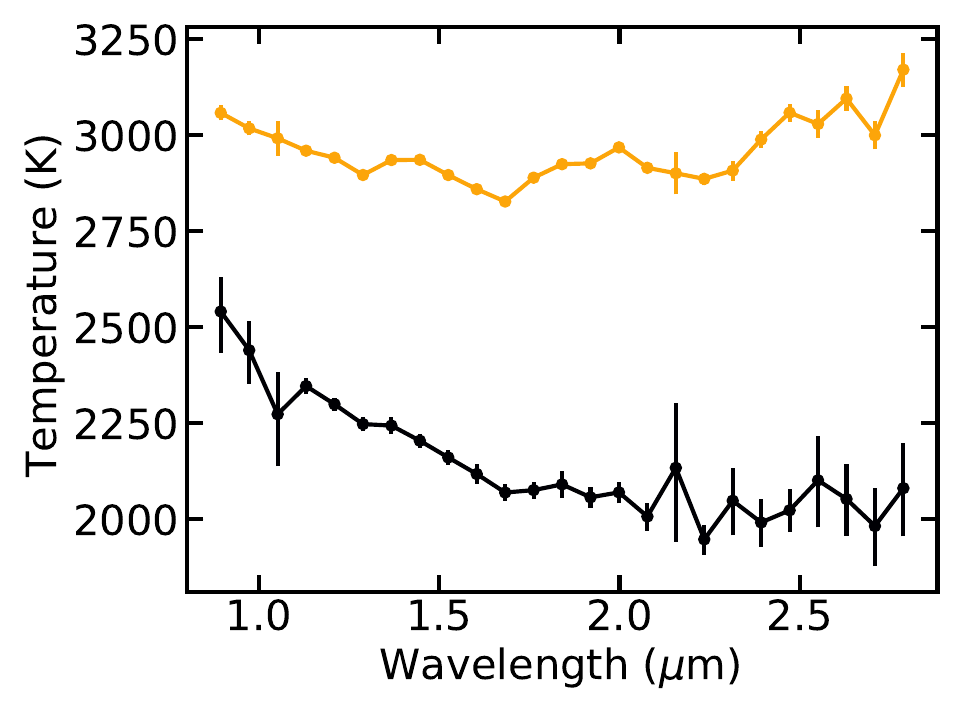}
    \includegraphics[width=0.35\linewidth]{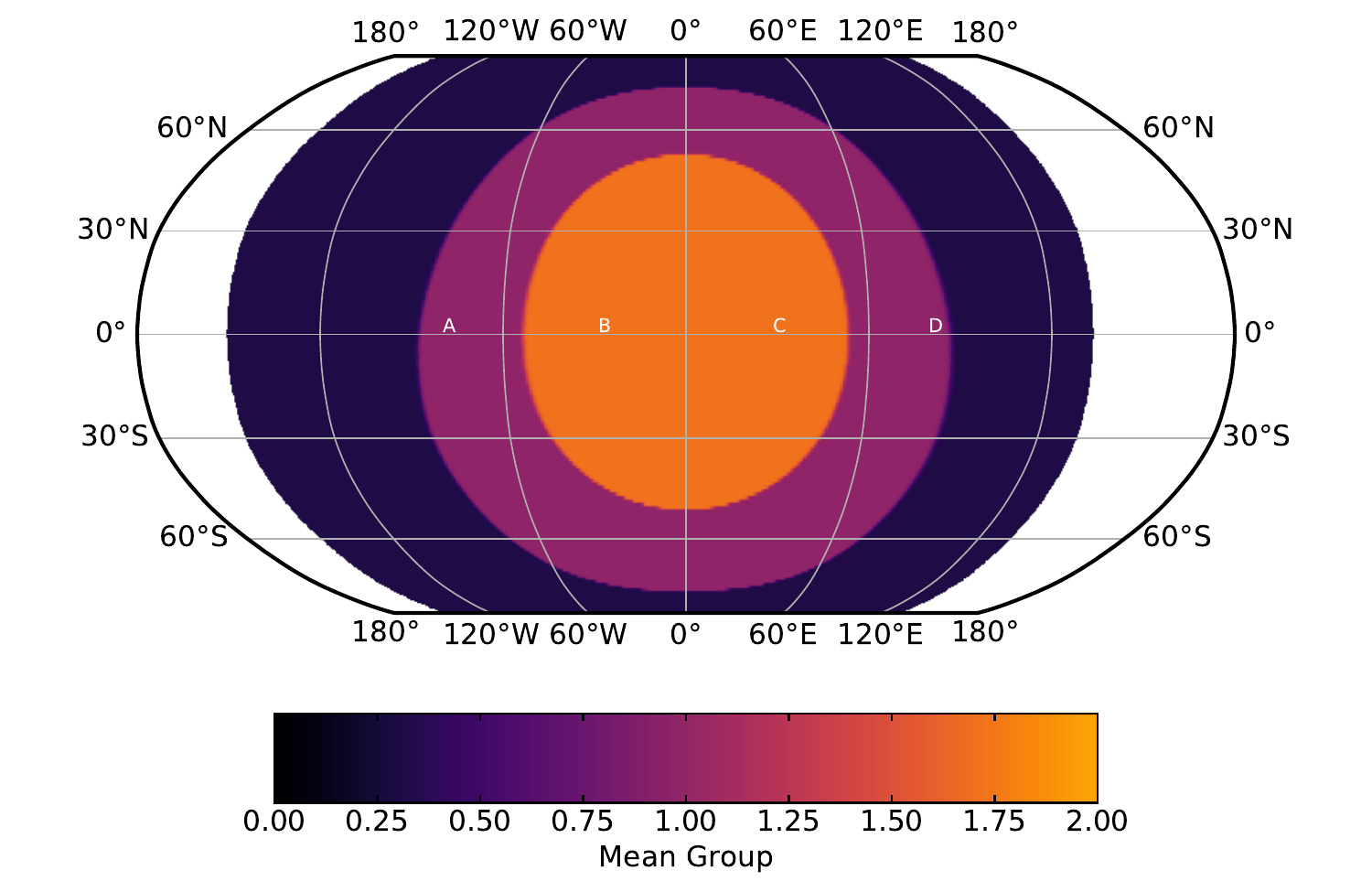}
    \includegraphics[width=0.35\linewidth]{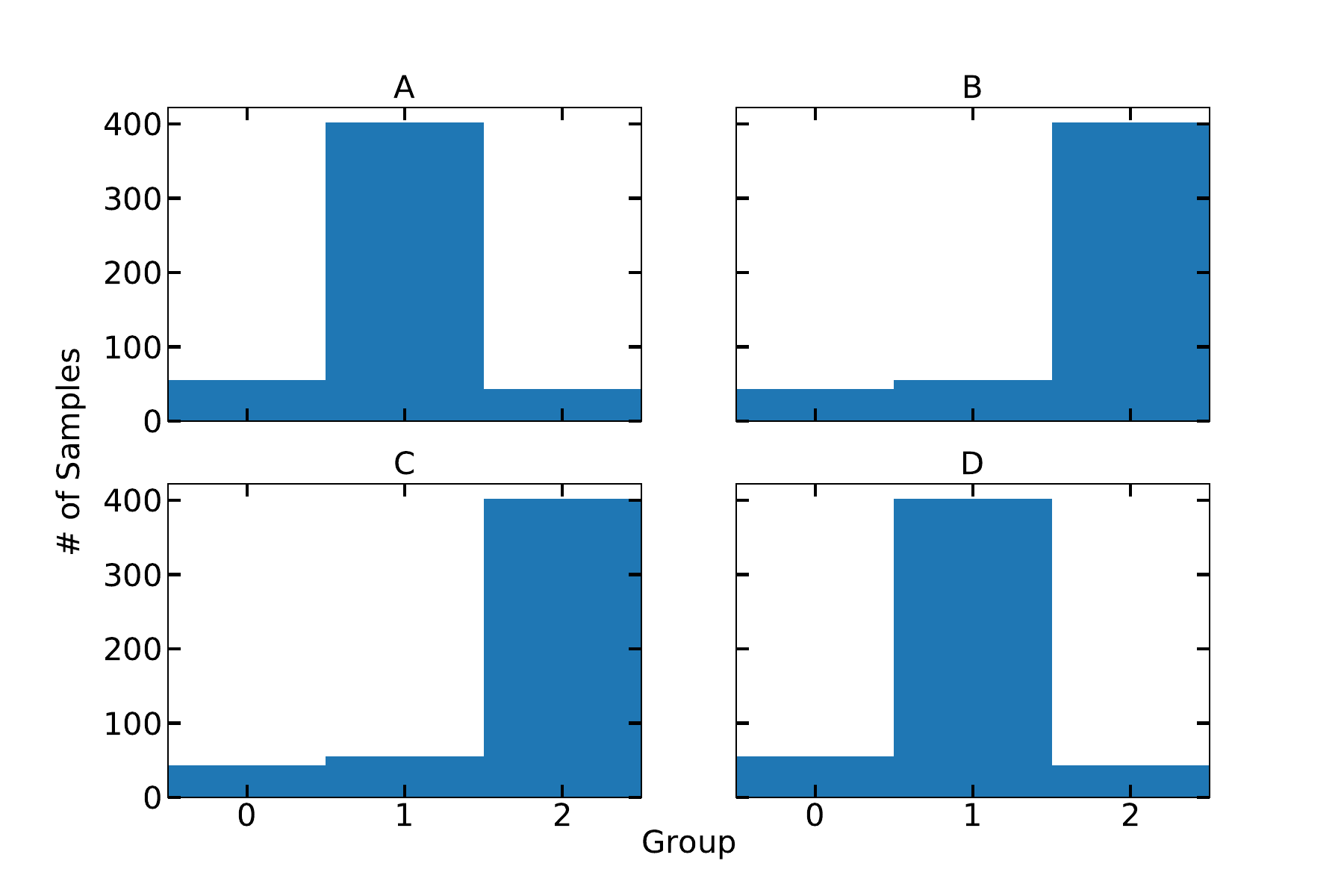}
    \includegraphics[width=0.28\linewidth]{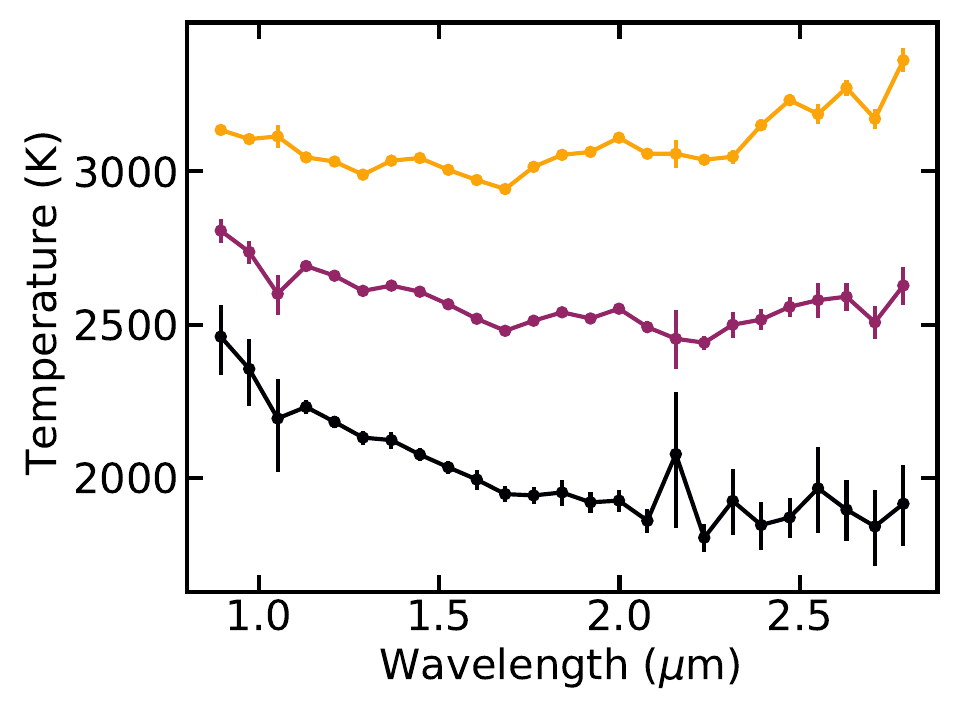}

    \includegraphics[width=0.35\linewidth]{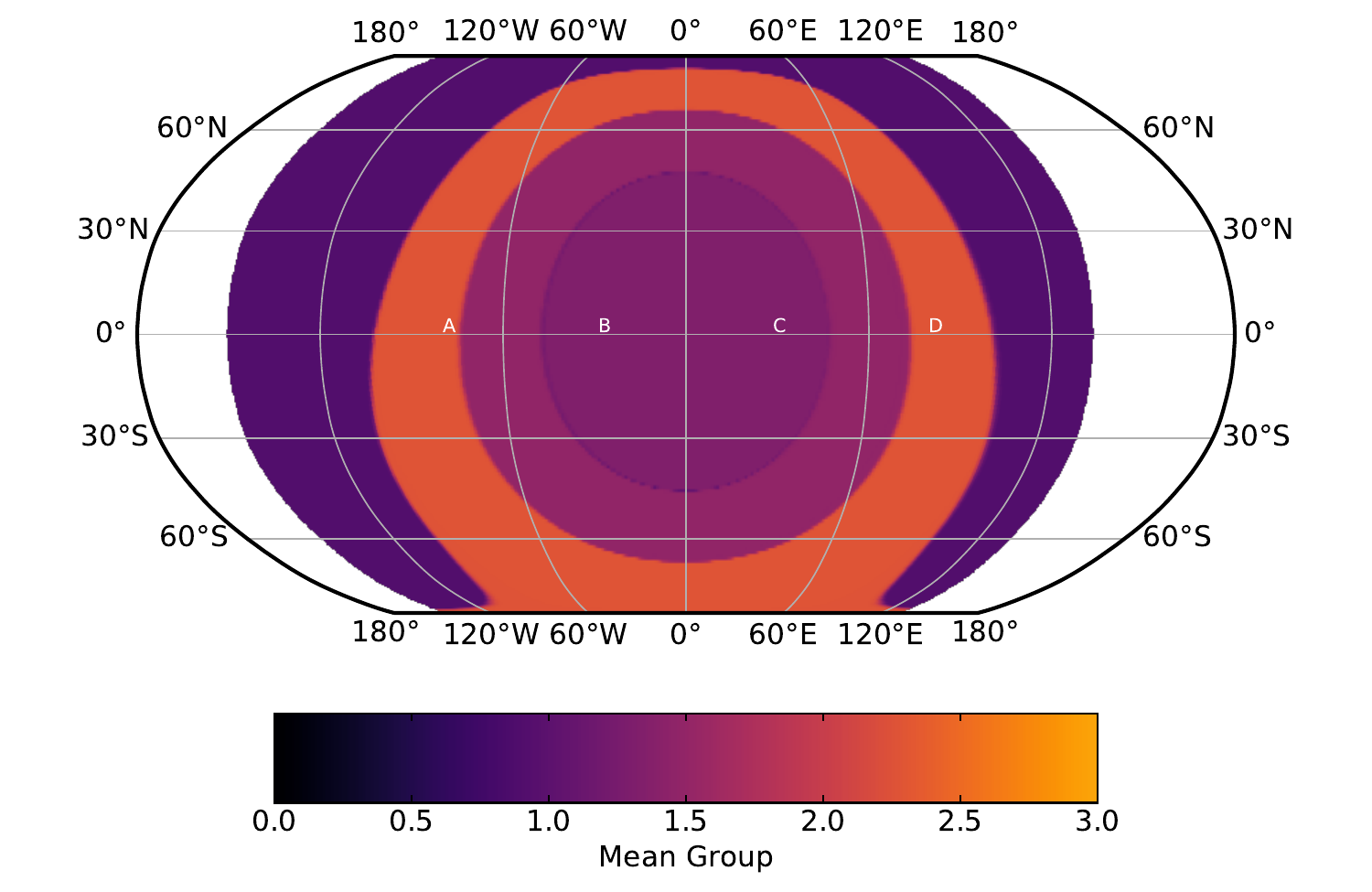}
    \includegraphics[width=0.35\linewidth]{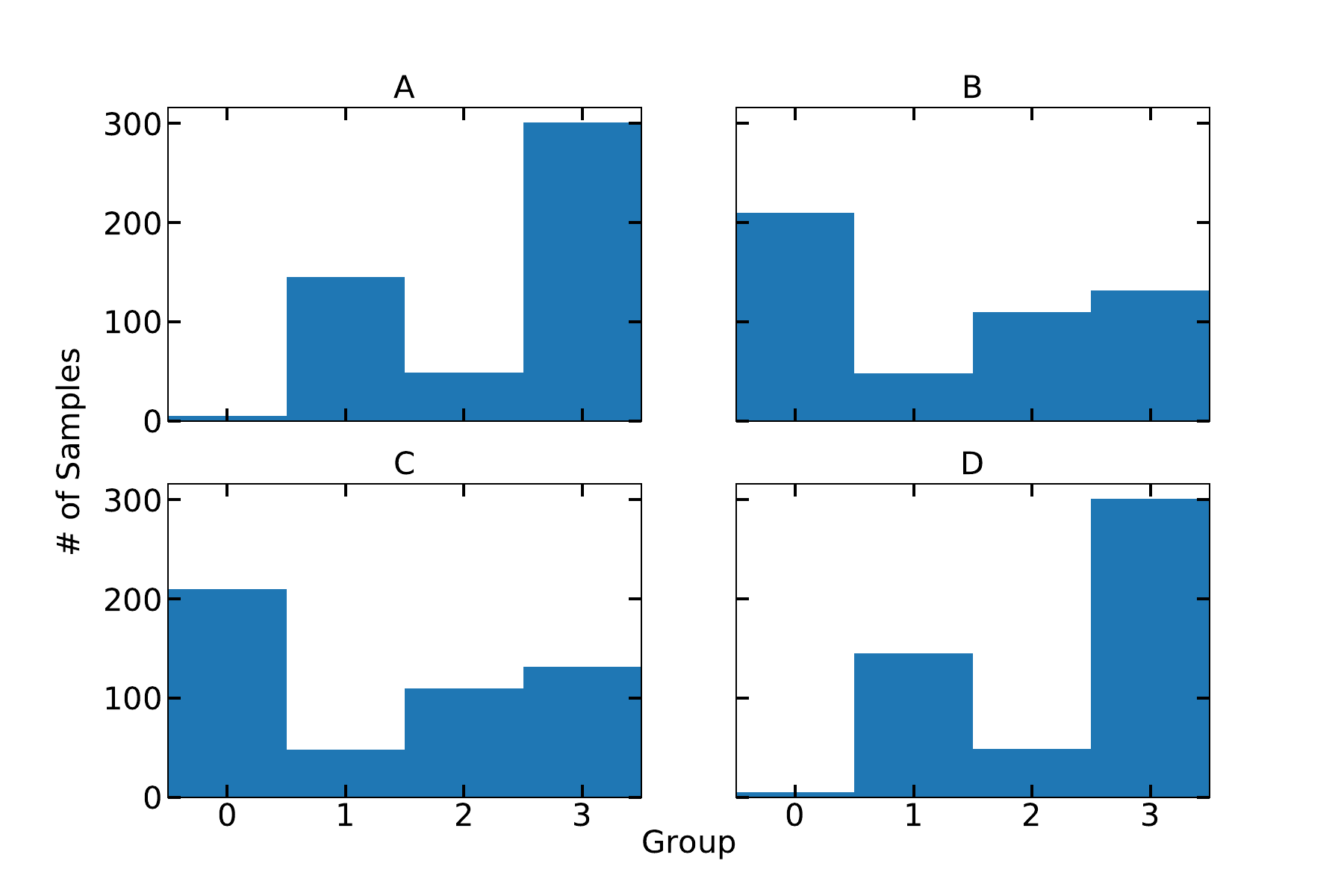}
    \includegraphics[width=0.28\linewidth]{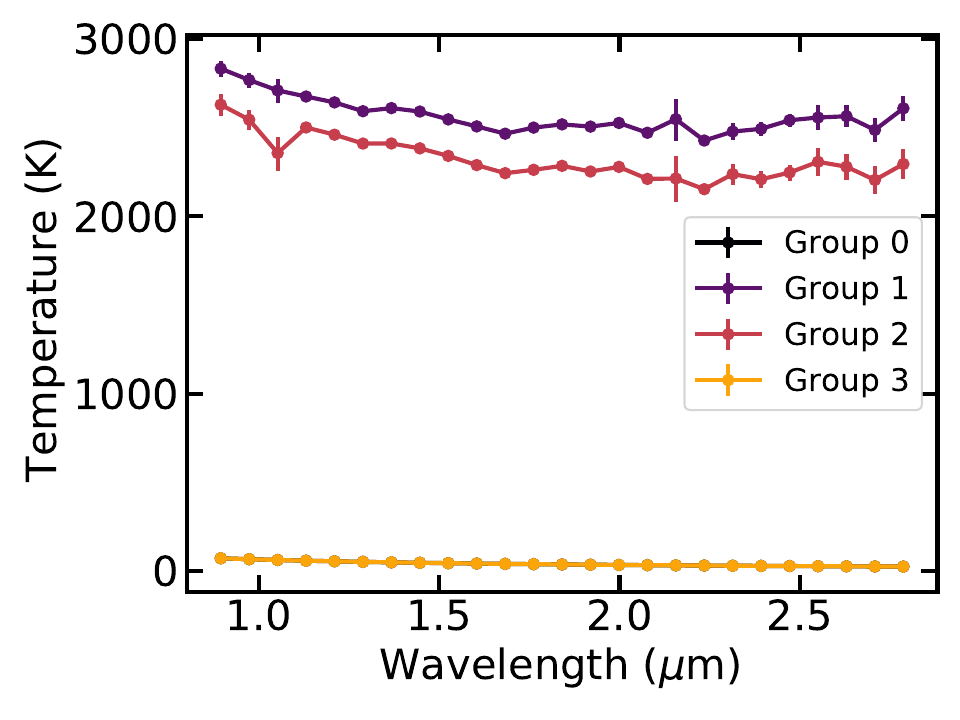}
    \caption{\sep Mean group (left), histograms of grouping across several MCMC samples (middle), and resulting eigenspectra (right) for \texttt{Eigenspectra} mapping fits using 2 (top), 3 (middle), and 4 (bottom) groups. For each set of plots, histograms are labelled by letters which are overplotted on the map in the latitude/longitude position from which they are drawn (positions were chosen both near and far from group edges). Groups 2 and 1 here correspond to the hotspot and ring groups discussed in the text. For 3 groups, the map shows a clear division between groups and all of the points show $\geq75$\% of points assigned to a single group. For 2 groups, there is a similarly clear division, and notably the hotspot and outer groups have quite similar spectra to the corresponding groups in the 3-group case. For 4 groups, the mean group map does not show a clear division of groups, and the histograms show that the same point is sorted into different groups depending on the posterior draw. Additionally, the resulting spectra are not distinct (the group 0 spectrum is identical to the group 4 spectrum, which is why it is not visible on the plot). Therefore, we used 3 groups for this fit.}
    \label{fig:eigen_group_histos}
\end{figure}

\begin{figure}[h]
    \centering
    \includegraphics[width=0.8\linewidth]{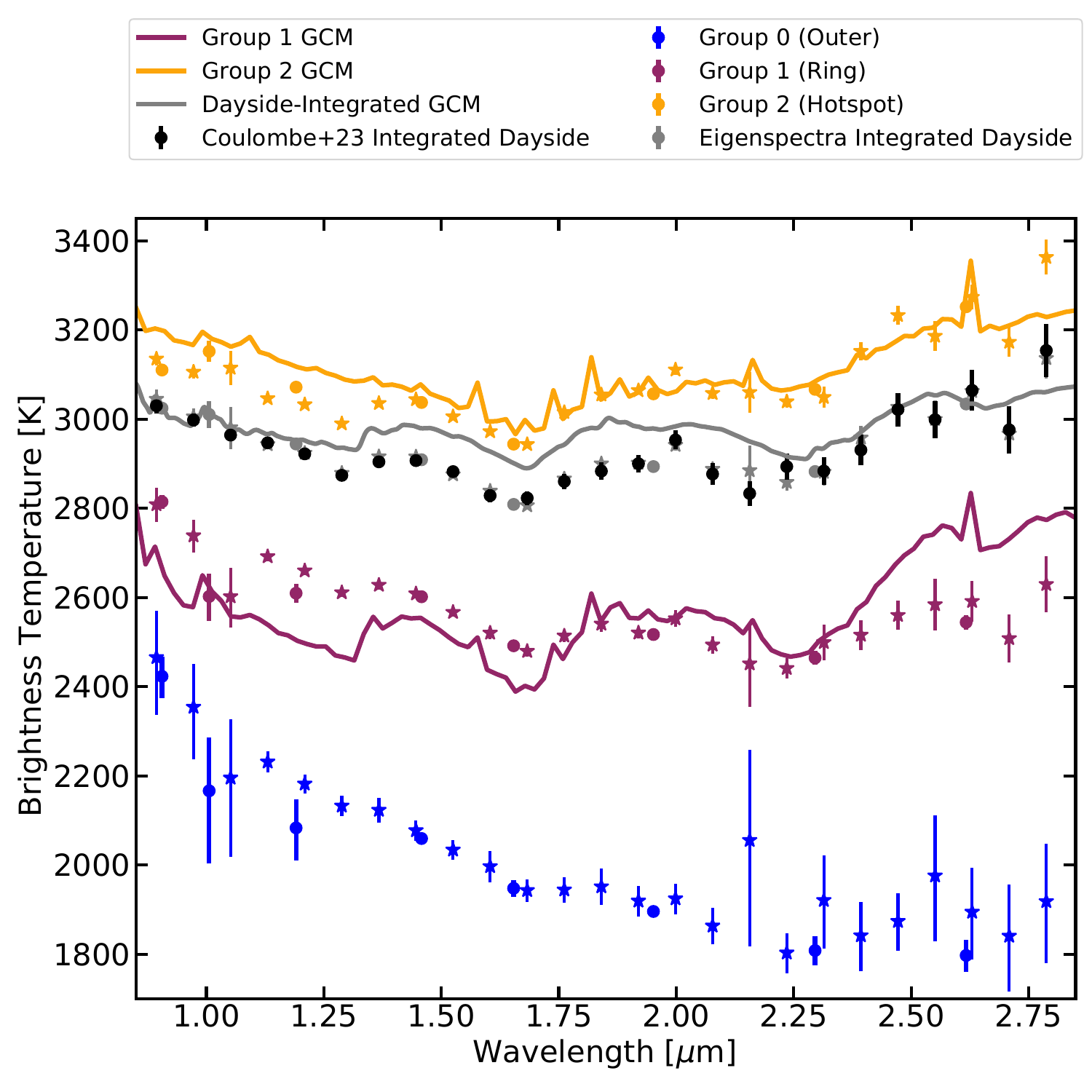}
    \caption{\sep Comparison of group spectra resulting from \texttt{Eigenspectra} and the full hemisphere-integrated dayside spectra on different resolution wavelength binning schemes. Black points show the full secondary eclipse spectrum from \cite{Coulombe2023} binned to the 25 wavelength bins discussed in the main text. Yellow, purple, and grey spectra show the output from \texttt{Eigenspectra} for the hotspot group, ring group, and the full hemisphere-integrated dayside spectrum. In all cases, stars show the higher resolution scheme with 25 wavelength bins discussed in the main text, while dots indicate the 8-wavelength bin scheme used in the methods for comparison to \texttt{ThERESA}. %, while stars show a higher-resolution scheme with 25 wavelength bins. 
    The hemisphere-integrated dayside spectrum from \texttt{Eigenspectra} closely matches the spectrum from \cite{Coulombe2023}, with all points within $1.2\sigma$ of the original spectrum, indicating that the fit is performing well. Additionally, there is little difference in the shapes of the spectra at higher vs. lower resolution. Lines show the dayside-integrated spectrum and grouped spectra from SPARC/MITgcm output. The predicted group spectra from the GCM provide a relatively good match to the Eigenspectra. For completeness, blue points show the output from \texttt{Eigenspectra} for the outer group, which was not analyzed fully due to its small contribution to the overall signal during the observation.}
    \label{fig:eigen_spectra_check}
\end{figure}

\begin{figure}[h]
    \centering
    \includegraphics[width=0.9\linewidth]{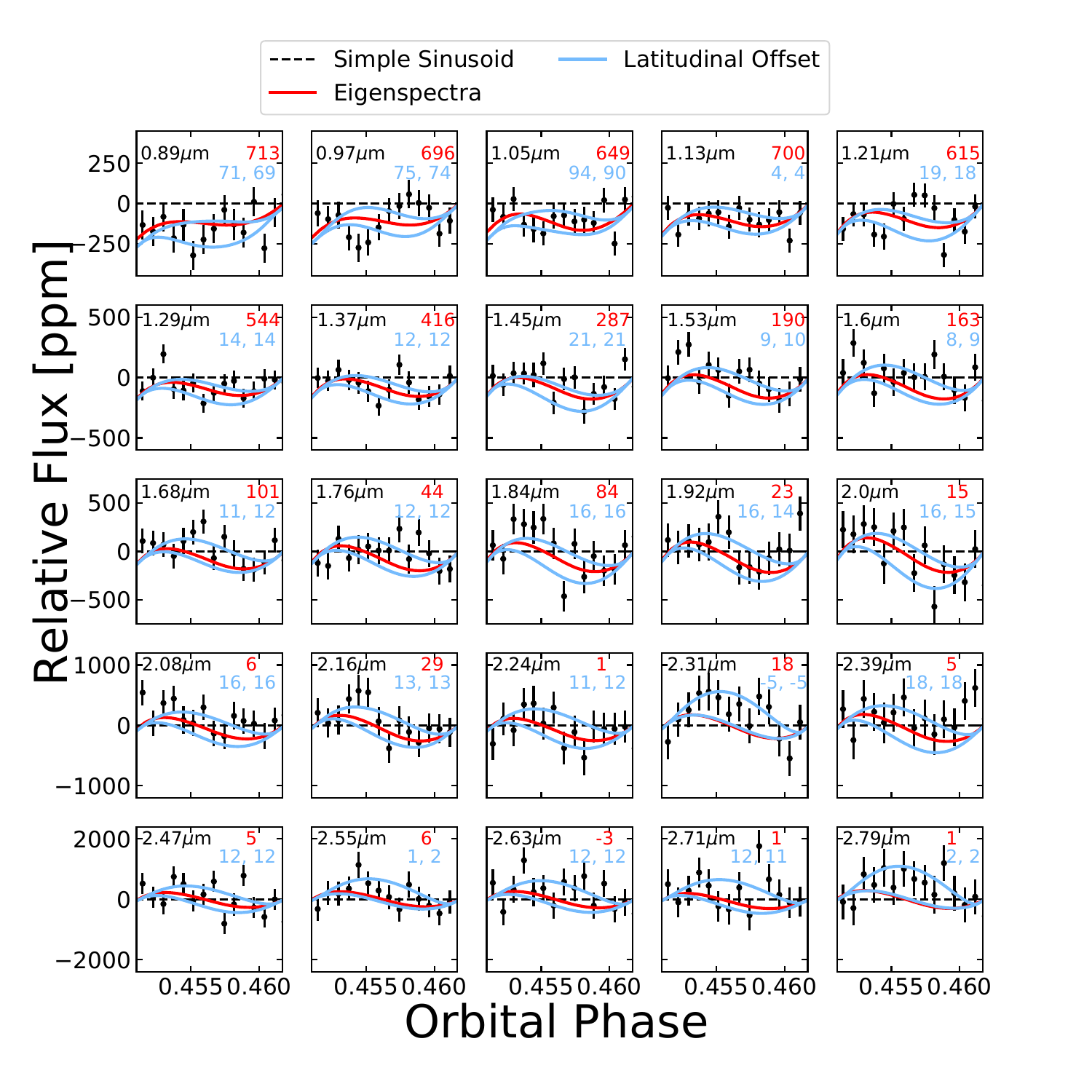}
    \caption{\sep Zoom-in on ingress for each of the 25 wavelength bins in the \texttt{Eigenspectra} fit. Panels show relative flux compared to the simple sinusoid fit described in the Methods (black lines), the real \texttt{Eigenspectra} fit (red), and the largest and smallest latitudinal offsets allowed for a fit with $\chi^{2} \leq10$ (blue lines). Black points with error bars show the difference in flux between the binned data and the simple sinusoid fit. Numbers in the upper right corners show the $\Delta \mathrm{BIC}$ for the fit to both ingress and egress between the simple sinusoid and \texttt{Eigenspectra} (red) and between the minimum and maximum latitudinal offset and \texttt{Eigenspectra} (blue). In all cases, a positive number indicates a lower BIC was achieved by the \texttt{Eigenspectra} fit. At more than half of the wavelengths fit, the \texttt{Eigenspectra} fit achieves a $\Delta \mathrm{BIC}\geq 10$ compared to all other fits, indicating a strong preference for \texttt{Eigenspectra}. The \texttt{Eigenspectra} fit is not as strongly preferred at longer wavelengths where the larger error bars naturally allow for a wider range of suitable fits to the data.}%The \texttt{Eigenspectra} fit provides a slight improvement over the simple sinusoid, with the reduced chi squared in each wavelength bin equal to or slightly lower than that for the sinusoid.}
    \label{fig:ingresszoom}
\end{figure}

\begin{figure}[h]
    \centering
    \includegraphics[width=0.9\linewidth]{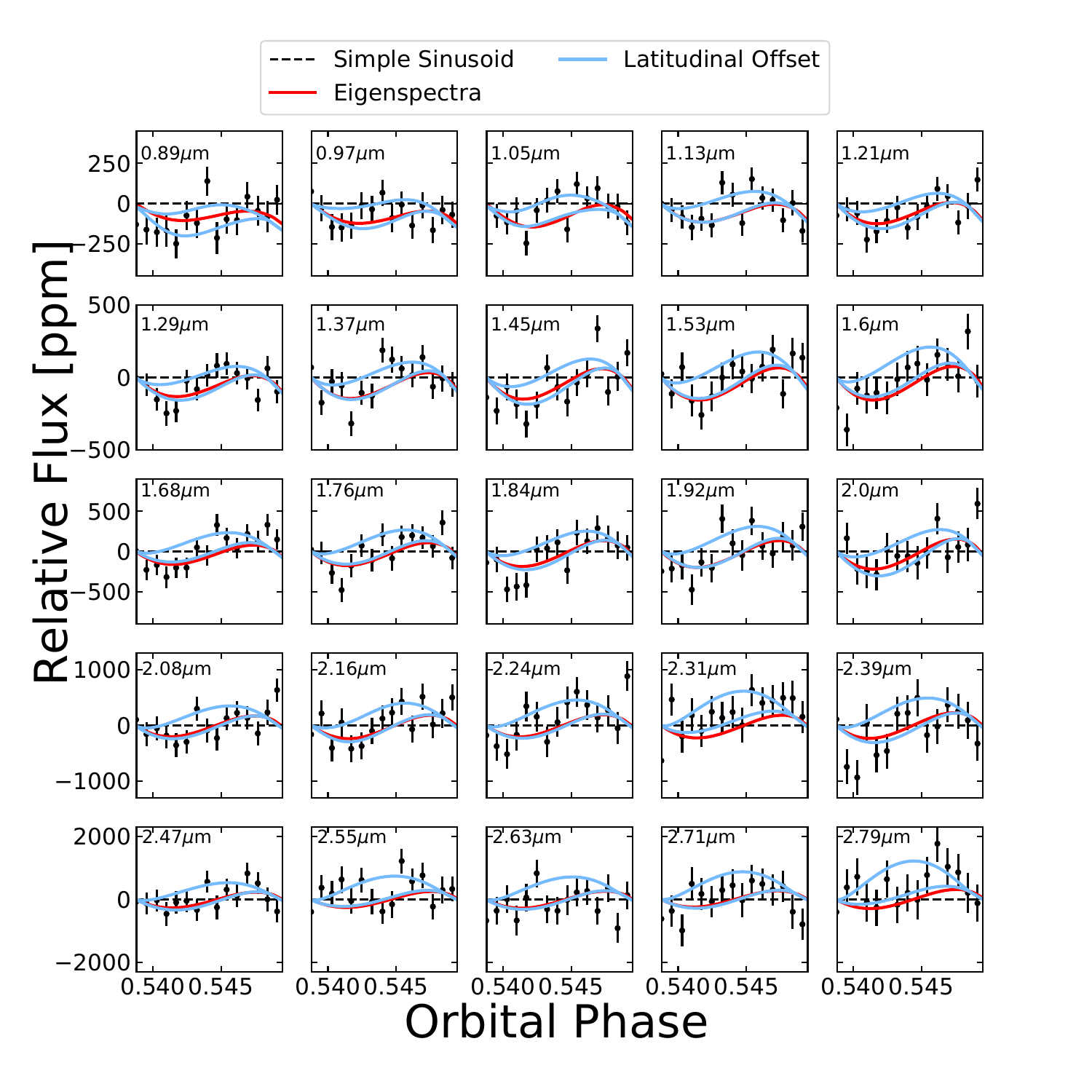}
    \caption{\sep Same as Supplementary Figure~\ref{fig:ingresszoom}, but showing the egress at each wavelength.}
    \label{fig:egresszoom}
\end{figure}

%     \caption{\sep {\color{red} \bf NOTE TO TEAM: FIGURE IS A WORK IN PROfGRESS, BUT WILL SHOW THE DIFFERENCE BETWEEN THE DATA AND A UNIFORM PLANET MODEL TO HIGHLIGHT THE ECLIPSE-MAPPING SIGNAL.}}
%     \label{fig:lcresiduals}
% \end{figure}

%     \caption{\sep Two-dimensional maps in units of brightness temperature for the 25 wavelength bin fit performed with the \texttt{Eigenspectra} method. As in Figure~\ref{fig:maps_8bin}, the transparency indicates the relative contribution to the overall observed flux at the point of maximum visibility, based on the angle between a given point on the map and the line of sight to the observer. Dotted black lines delineate the three regions identified by the \texttt{Eigenspectra} method.}
%     \label{fig:maps25bin}
% \end{figure}

%     \caption{\sep Light curve fits for the 25 wavelength bin fit performed with the \texttt{Eigenspectra} method. As in Figure~\ref{fig:maps_8bin}, black points with error bars indicate wavelength-binned, systematics-corrected data from Coulombe et al. (2023)\cite{Coulombe2023}, and red lines show best fits from the \texttt{Eigenspectra} method.}
%     \label{fig:lc25bin}
% \end{figure}

\begin{table}
    \centering
    \begin{tabular}{c | c c c | c}
        Wavelength [$\mu$m] & Hotspot [K] & Ring [K] & Outer [K] & Hotspot Offset [$\degree$] \\
        \hline
        $0.85-0.93$ & $3135\pm15$ & $2809^{+40}_{-42}$ & $2472^{+114}_{-144}$ & $1.5^{+1}_{-3}$ \\
        $0.93-1.01$ & $3106\pm14$ & $2738^{+34}_{-35}$ & $2345^{+89}_{-107}$ & $-5.5\pm2$ \\
        $1.01-1.09$ & $3114^{+36}_{-37}$ & $2601^{+62}_{-67}$ & $2193^{+133}_{-182}$ & $-0.5^{+2}_{-1}$ \\
        $1.09-1.17$ & $3046\pm6$ & $2692\pm11$ & $2233^{+25}_{-26}$ & $3.5\pm1$ \\
        $1.17-1.25$ & $3032\pm5$ & $2660\pm9$ & $2183^{+21}_{-22}$ & $-4.5\pm3$ \\
        $1.25-1.33$ & $2989^{+5}_{-6}$ & $2611\pm10$ & $2133^{+23}_{-24}$ & $-2.5^{+1}_{-0}$ \\
        $1.33-1.41$ & $3035\pm6$ & $2628\pm11$ & $2126^{+29}_{-30}$ & $-1.5^{+1}_{-0}$ \\
        $1.41-1.49$ & $3044\pm6$ & $2608\pm10$ & $2077\pm22$ & $2.5^{+0}_{-1}$ \\
        $1.49-1.56$ & $3005\pm6$ & $2566\pm11$ & $2033^{+22}_{-23}$ & $3.5^{+2}_{-4}$ \\
        $1.56-1.64$ & $2972\pm8$ & $2520\pm13$ & $1994^{+32}_{-34}$ & $-1.5\pm1$ \\
        $1.64-1.72$ & $2943\pm8$ & $2480\pm13$ & $1946^{+28}_{-29}$ & $1.5^{+0}_{-2}$ \\
        $1.72-1.80$ & $3015\pm8$ & $2514\pm14$ & $1944^{+29}_{-30}$ & $-0.5^{+2}_{-1}$ \\
        $1.80-1.88$ & $3054\pm10$ & $2542\pm18$ & $1956^{+42}_{-44}$ & $-3.5^{+1}_{-2}$ \\
        $1.88-1.96$ & $3064\pm8$ & $2521\pm13$ & $1922^{+34}_{-36}$ & $-1.5\pm1$ \\
        $1.96-2.04$ & $3110\pm11$ & $2553\pm8$ & $1926^{+33}_{-34}$ & $-1.5\pm1$ \\
        $2.04-2.12$ & $3058\pm12$ & $2493\pm20$ & $1863^{+41}_{-42}$ & $-5.5^{+2}_{-1}$ \\
        $2.12-2.20$ & $3057^{+44}_{-45}$ & $2456^{+91}_{-94}$ & $2079^{+195}_{-228}$ & $4.5^{+2}_{-1}$ \\
        $2.20-2.27$ & $3038\pm15$ & $2441\pm24$ & $1804^{+45}_{-47}$ & $-1.5^{+1}_{-0}$ \\
        $2.27-2.35$ & $3047\pm24$ & $2502\pm42$ & $1930^{+104}_{-113}$ & $-0.5^{+0}_{-1}$ \\
        $2.35-2.43$ & $3151\pm21$ & $2517^{+34}_{-35}$ & $1845^{+80}_{-86}$ & $2.5\pm1$ \\
        $2.43-2.51$ & $3232\pm20$ & $2559\pm32$ & $1872^{+62}_{-65}$ & $-1.5^{+1}_{-3}$ \\
        $2.51-2.59$ & $3187\pm32$ & $2581^{+55}_{-56}$ & $1967^{+127}_{-138}$ & $1.5^{+1}_{-0}$ \\
        $2.59-2.67$ & $3271\pm29$ & $2595^{+48}_{-49}$ & $1905^{+107}_{-115}$ & $-0.5^{+4}_{-2}$ \\
        $2.67-2.75$ & $3170^{+33}_{-34}$ & $2512^{+55}_{-56}$ & $1853^{+121}_{-131}$ & $7.5\pm1$ \\
        $2.75-2.83$ & $3362\pm40$ & $2631^{+64}_{-65}$ & $1924^{+132}_{-143}$ & $2.5^{+2}_{-1}$ \\
    \end{tabular}
    \caption{\sep The three group spectra extracted from \texttt{Eigenspectra}, using 25 bins evenly spaced in wavelength, in units of brightness temperature. We also list the hotspot offset and error bar at each wavelength. Note that all offsets are half-integer values and there are some wavelengths where the 1$\sigma$ error bar appears to be zero. This is because of the longitude grid cell spacing ($1\degree$), which naturally only allows for half-integer values and integer error bars.}
    \label{tab:eigenspec25}
\end{table}

\begin{table}
    \centering
    \begin{tabular}{c | c c c | c}
        Wavelength [$\mu$m] & Hotspot [K] & Ring [K] & Outer [K] & Hotspot Offset [$\degree$] \\
        \hline
        $0.86-0.96$ & $3110\pm8$ & $2814\pm16$ & $2423^{+49}_{-45}$ & $1.5^{+1}_{-2}$ \\
        $0.96 - 1.06$ & $3152\pm24$ & $2602^{+54}_{-51}$ & $2166^{+163}_{-120}$ & $2.5^{+1}_{-1}$ \\
        $1.06-1.33$ & $3072\pm9$ & $2609\pm21$ & $2083^{+72}_{-64}$ & $1.5^{+0}_{-2}$ \\
        $1.33-1.59$ & $3037\pm3$ & $2601\pm5$ & $2059\pm10$ & $-1.5^{+1}_{-0}$ \\
        $1.59-1.72$ & $2944\pm6$ & $2492\pm10$ & $1948\pm19$ & $-0.5^{+0}_{-1}$ \\
        $1.72-2.18$ & $3057\pm4$ & $2517\pm6$ & $1896\pm9$ & $-1.5^{+0}_{-1}$ \\
        $2.18 - 2.41$ & $3067\pm9$ & $2464\pm15$ & $1808^{+33}_{-32}$ & $1.5^{+1}_{-2}$ \\
        $2.41 - 2.83$ & $3252\pm11$ & $2544\pm17$ & $1797^{+36}_{-35}$ & $-0.5^{+2}_{-1}$ \\
    \end{tabular}
    \caption{\sep Same as Supplementary Table~\ref{tab:eigenspec25}, but for the 8 optimally-chosen wavelength bins.}
    \label{tab:eigenspectra}
\end{table}

\begin{figure}[h]
    \centering
    \includegraphics[width=\linewidth]{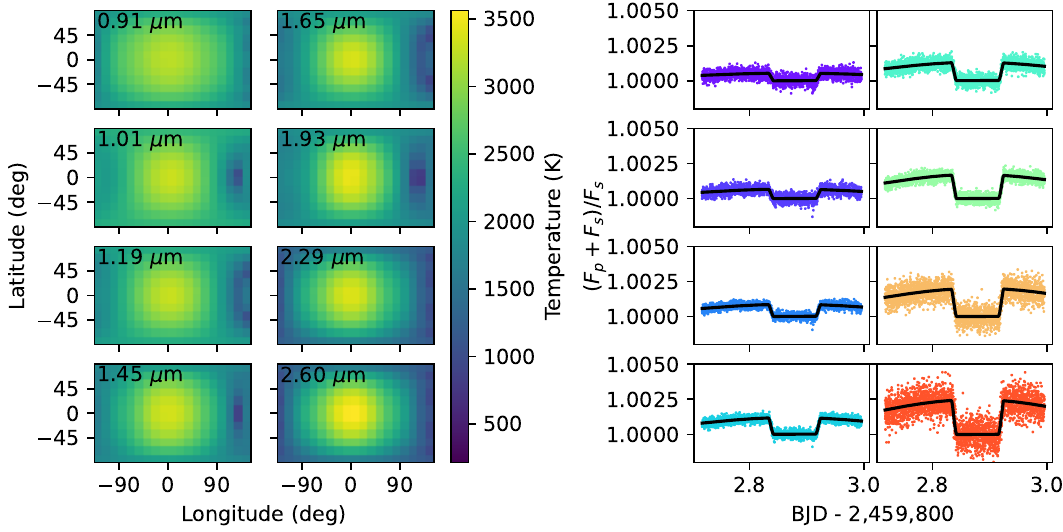}
    \caption{\sep 2D brightness temperature maps from \texttt{ThERESA}, at each of the 8-wavelength bins, and the corresponding light-curve fits. The 2D mapping method is the same as used in \texttt{Eigenspectra}, and the resulting maps are consistent with those in Figure \ref{fig:maps_8bin}.}
    \label{fig:theresa2d}
\end{figure}

\begin{figure}[h]
    \includegraphics[width=\textwidth]{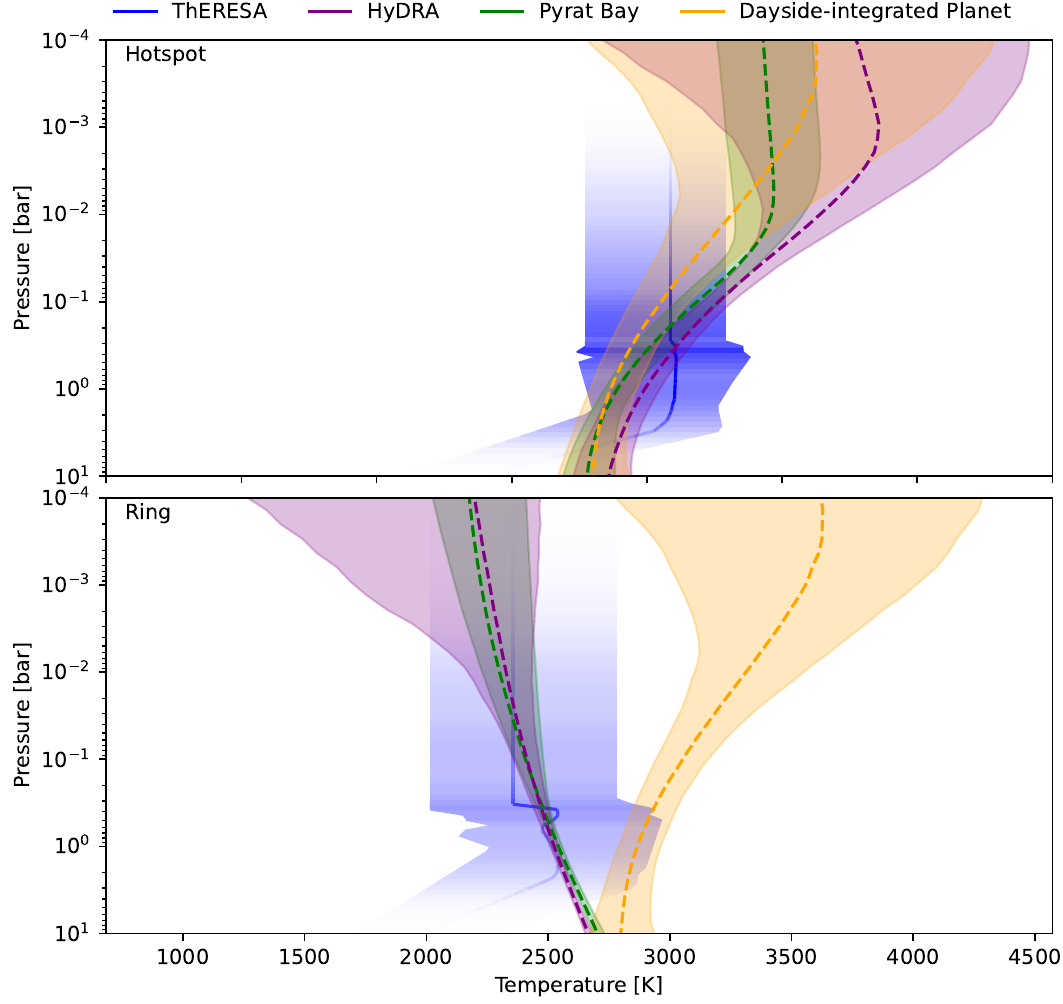}
    \caption{\sep The range of vertical temperature profiles and median (by pressure layer) profile from the best-fitting \texttt{ThERESA} 3D temperature map (blue), grouped into the \texttt{Eigenspectra} regions and compared against the profiles retrieved from \texttt{Eigenspectra} (purple, green) and the full-planet spectrum (orange). The transparency of the range of vertical temperature profiles and the median profile have been scaled by the contribution function at each location, showing which vertical locations are probed by the data. The jagged temperature profiles are caused by the linear interpolation (see text). At the hotspot, in the pressures primarily probed by our observation ($\sim0.03 - 3$ bar), \texttt{ThERESA} finds a mix of inverted thermal profiles near the substellar point (upper bound of the blue region) and more isothermal profiles further from the center of the hotspot. In the ring, the temperature profiles are largely non-inverted at the pressures we probe. In both cases, we see general agreement with the 1D retrievals on the \texttt{Eigenspectra}.}
    \label{fig:theresa3dtp}
\end{figure}

\begin{figure}[h]
    \centering
    \includegraphics[width=0.84\linewidth]{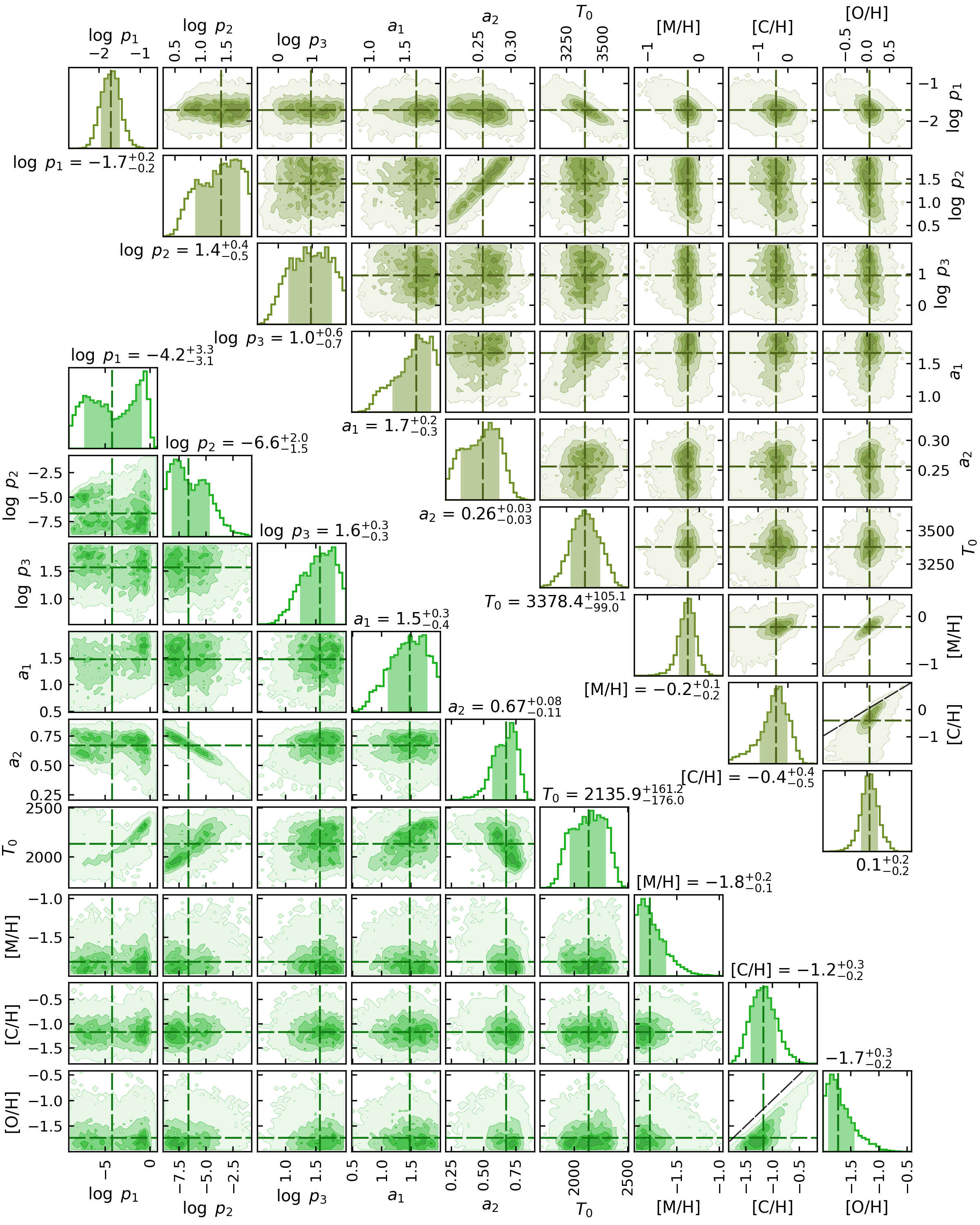}
    \caption{\sep Pairs plot showing posterior distributions for parameters in the {\pyratbay} retrieval of the hotspot group (top right) and ring group (bottom left). The first six parameters ($\log p_1$, $\log p_2$, $\log p_3$, $a_1$, $a_2$, $T_0$) determine the P-T profile model\cite{MadhusudhanSeager2009apjRetrieval}. [C/H] and [O/H] are the carbon and oxygen elemental abundances (respectively) relative to solar abundances. [M/H] is a catch-all parameter to scale the abundance of all other metals relative to solar. Off-diagonal plots show 2D posterior probabilities for pairs of parameters, with probability densities shaded in green. On-diagonal plots show marginalized posterior probability distributions for each parameter. Quoted values denote the median and central 68\% fraction of the marginal posterior distributions. The black dashed lines trace the constant C/O curve equal to the solar value.}
    \label{fig:hotspot_pairs_pyratbay}
\end{figure}

\begin{figure}[h]
    \centering
    \includegraphics[width=0.95\linewidth]{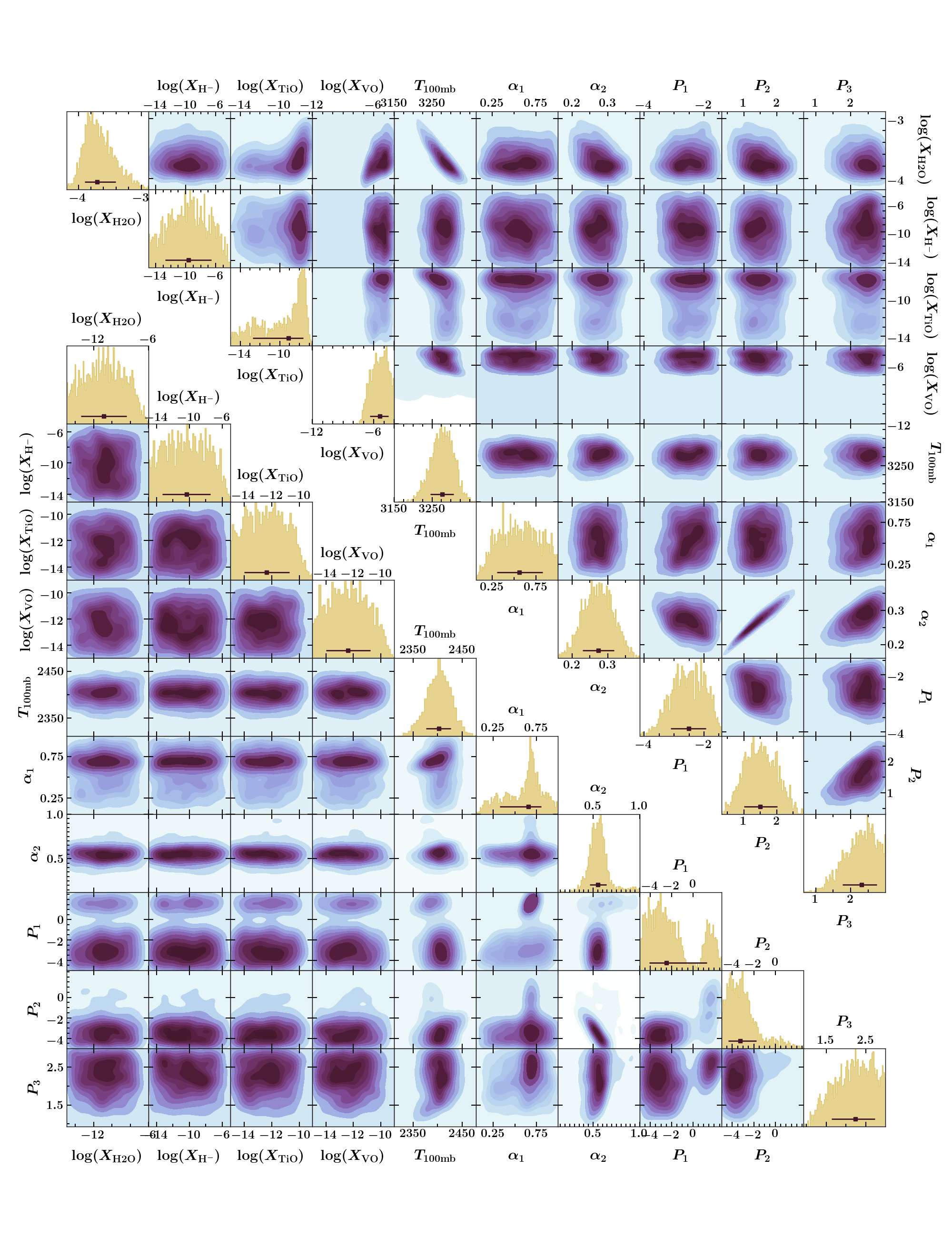}
    \vspace{-1.4cm}
    \caption{\sep Similar to Supplementary Figure~\ref{fig:hotspot_pairs_pyratbay}, but showing pairs plots for the \texttt{HyDRA} retrievals of the hotspot group (upper right) and the ring group standard model (lower left). log($X_i$) are the log mixing ratios of species $i$, T$_\mathrm{100mb}$ is the temperature at 100~mbar and $\alpha_1$, $\alpha_2$, P$_1$, P$_2$ are P-T profile parameters as described in ref\cite{MadhusudhanSeager2009apjRetrieval}.}
    \label{fig:pairs_hydra}
\end{figure}

\begin{figure}[h]
    \centering
    \includegraphics[width=\linewidth]{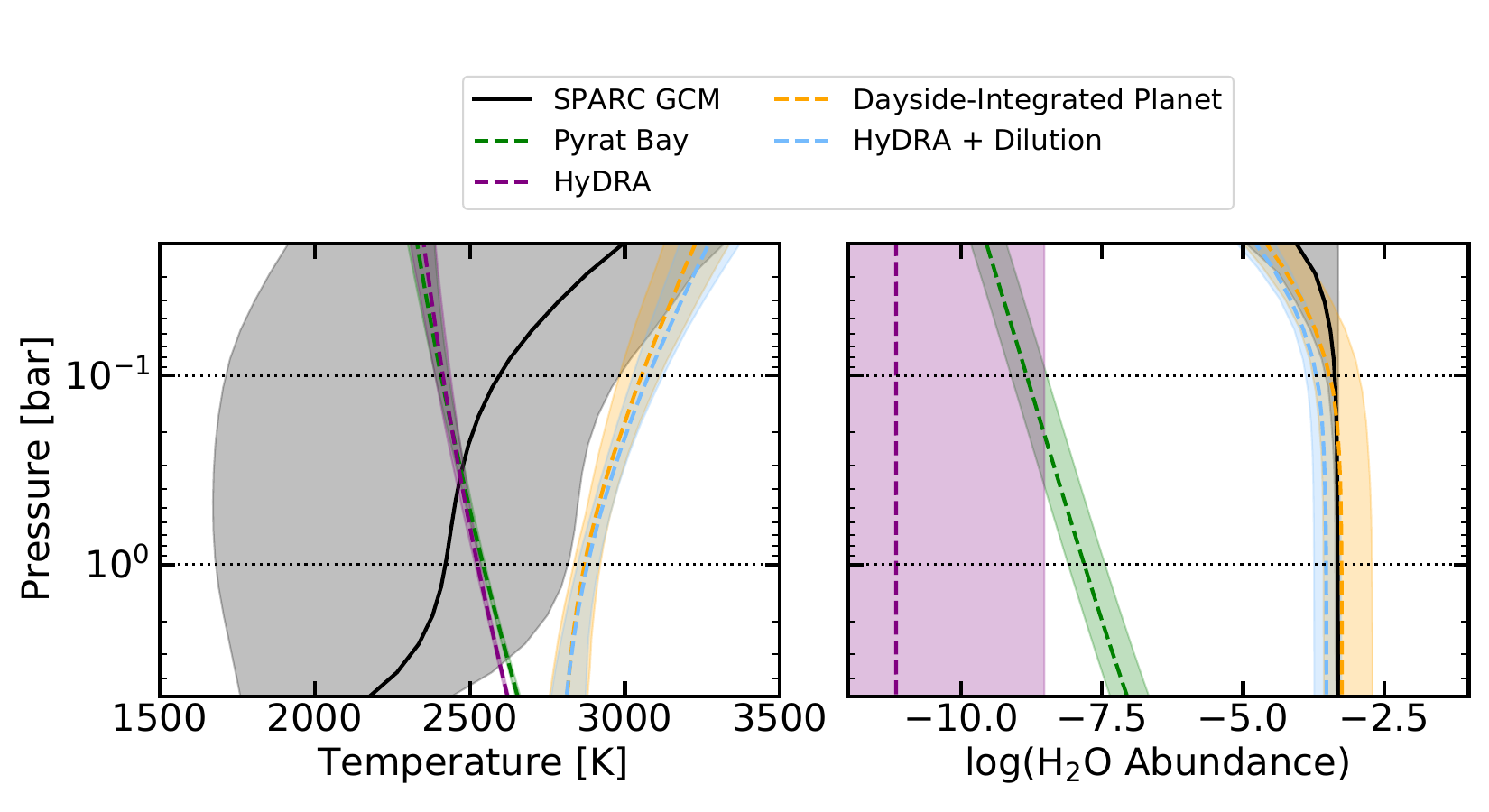}
    \caption{\sep Retrieved T-P profiles (left) and H$_{2}$O abundance (right) for the \texttt{Eigenspectra} ring group. In all plots, purple and green lines show the standard retrievals following the same set-up as for the hotspot group using \texttt{HyDRA} and {\pyratbay}, respectively, and yellow lines show the retrieval on the full dayside spectrum\cite{Coulombe2023}, with shading showing $1\sigma$ confidence intervals. Black solid lines show average profiles in the ring group region from a SPARC/MIT GCM, and black shaded regions show the full range of per-point GCM profiles in that region. We also show one additional retrieval run with \texttt{HyDRA} - the standard model with the addition of a dilution parameter (blue lines and $1\sigma$ shaded region). %and a blackbody model with a dilution parameter (tan lines and $1\sigma$ shaded region). 
    The mean posterior value of the dilution parameter is 0.64, with a 2$\sigma$ credible region of $0.59 - 0.69$.
    Black dotted lines indicate the approximate extent of pressures probed across all models. Changing the model set-up drastically changes both the T-P profile and the retrieved abundances. We discuss several potential explanations for these changes and future directions for research to better understand these discrepancies in the Methods.}
    \label{fig:retrieve_ring}
\end{figure}

%    \caption{\sep Same as Figure~\ref{fig:hotspot_pairs_pyratbay}, but for the {\pyratbay} retrieval of the ring group.}
%    \label{fig:ring_pairs_pyratbay}
%\end{figure}

\begin{figure}[h]
    \centering
    \includegraphics[width=\linewidth]{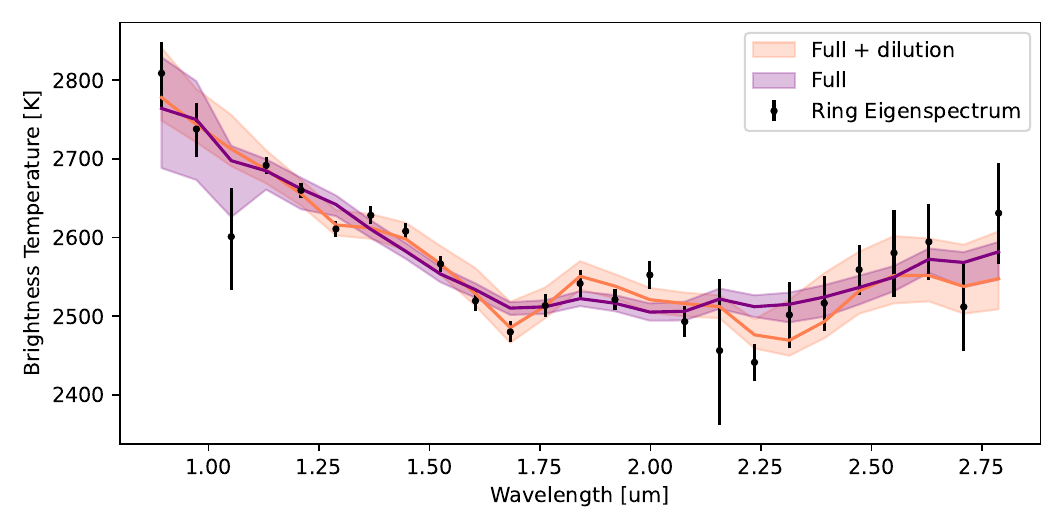}
    \caption{\sep Comparison of resulting spectra from different fits to the \texttt{Eigenspectra} ring spectrum (black points with error bars) with \texttt{HyDRA}. Purple and orange lines show the standard model and the standard model with the addition of a dilution parameter, respectively. Shaded areas indicate 95.45\% credible regions. The standard model with dilution provides the best fit to the data and matches the slight water emission features seen by eye near $1.4$ and $1.9$ $\mu$m, but as discussed in the Methods, the \texttt{Eigenspectra} method is designed to eliminate the need for a dilution parameter. Therefore, the preference for the standard+dilution model is likely obscuring some unaccounted-for physical or geometric effects.}
    \label{fig:ring_spectra}
\end{figure}

\clearpage
\subsection{References}

\bibliography{main}

%\end{supplementary}
\end{document}